\documentclass[english,12pt]{article}
\usepackage[T1]{fontenc}
\usepackage[latin9]{inputenc}
\usepackage{geometry}
\usepackage{bm}
\usepackage{amsmath}
\usepackage{amsthm}
\usepackage{amssymb}
\usepackage{graphicx}
\usepackage[authoryear]{natbib}
\usepackage{setspace}
\usepackage{xcolor}

\makeatletter

\providecommand{\tabularnewline}{\\}

\newcommand{\lyxaddress}[1]{
	\par {\raggedright #1
	\vspace{1.4em}
	\noindent\par}
}
\theoremstyle{plain}
\newtheorem{thm}{\protect\theoremname}
\theoremstyle{plain}
\newtheorem{cor}{\protect\corollaryname}

\usepackage{url}

\makeatother

\usepackage{babel}
\providecommand{\corollaryname}{Corollary}
\providecommand{\theoremname}{Theorem}

\begin{document}
\title{Order selection with confidence for finite mixture models}
\author{Hien D Nguyen$^{1*}$ \and Daniel Fryer$^{1}$ \and Geoffrey J McLachlan$^{1}$}
\maketitle

\lyxaddress{
$^{1}$School of Mathematics and Physics, University of Queensland,
St. Lucia, Australia. \\
$^{*}$Corresponding author: \url{h.nguyen5@latrobe.edu.au}.}
\begin{abstract}
The determination of the number of mixture components (the order)
of a finite mixture model has been an enduring problem in statistical
inference. We prove that the closed testing principle leads to
a sequential testing procedure (STP) that allows for confidence
statements to be made regarding the order of a finite mixture model.
We construct finite sample tests, via data splitting and data swapping,
for use in the STP, and we prove that such tests are consistent against
fixed alternatives. Simulation studies {and real data examples} are used to demonstrate
the performance of the finite sample tests-based STP, yielding practical recommendations {of their use as confidence estimators in combination with point estimates such as the Akaike information or Bayesian information criteria}. In addition, we demonstrate that a modification
of the STP yields a method that consistently selects the order of
a finite mixture model, in the asymptotic sense. Our STP is not only applicable
for order selection of finite mixture models, but is also useful for making
confidence statements regarding any sequence of nested models.
\end{abstract}


\section{\label{sec:Introduction}Introduction}

Let $\bm{X}\in\mathbb{X}$ be a random variable. Let $\mathcal{K}\left(\mathbb{X}\right)$
be a class of probability density functions (PDFs), defined on the
set $\mathbb{X}$, which we shall refer to as components. We say that
$\bm{X}$ arises from a $g$ component mixture model of class $\mathcal{K}$
if the PDF $f_{0}$ of $\bm{X}$ belongs in the convex class
\[
\mathcal{M}_{g}\left(\mathbb{X}\right)=\left\{ f\left(\bm{x}\right):f\left(\bm{x}\right)=\sum_{z=1}^{g}\pi_{z}f_{z}\left(\bm{x}\right);\pi_{z}\ge0,\sum_{z=1}^{g}\pi_{z}=1,f_{z}\in\mathcal{K}\left(\mathbb{X}\right),z\in\left[g\right]\right\} \text{,}
\]
where $g\in\mathbb{N}$ and $\left[g\right]=\left\{ 1,\dots,g\right\} $.

Suppose that we observe an independent and identically distributed
(IID) sample sequence of data $\mathbf{X}_{n}=\left(\bm{X}_{i}\right)_{i=1}^{n}$,
where each $\bm{X}_{i}$ has the same data generating process (DGP)
as $\bm{X}$, which is unknown. Under the assumption that $f_{0}\in\mathcal{M}_{g_{0}}\left(\mathbb{X}\right)$
for some $g_{0}\in\mathbb{N}$, we wish to use the data $\mathbf{X}_{n}$
in order to determine the possible values of $g_{0}$. This problem
is generally referred to as order selection in the mixture modeling
literature, and reviews regarding the problem can be found in \citet[Ch. 6]{McLachlan2000} and
\citet{McLachlan2014}, for example.

Notice that the sequence $\left(\mathcal{M}_{g}\right)_{g=1}^{\infty}$
is nested, in the sense that $\mathcal{M}_{g}\subset\mathcal{M}_{g+1}$,
for each $g$, and that $g_{0}\in\mathbb{N}$ is equivalent to $f_{0}\in\mathcal{M}=\bigcup_{g=1}^{\infty}\mathcal{M}_{g}$.
We shall write the null hypothesis that $f_{0}\in\mathcal{M}_{g}$
(or equivalently, $g_{0}\le g$) as $\text{H}_{g}$, and we assume
that we have available a $p\text{-value}$ $P_{g}\left(\mathbf{X}_{n}\right)$ for each hypothesis, and that $P_{g}\left(\mathbf{X}_{n}\right)$ correctly controls the size of the
hypothesis test, in the sense that
\begin{equation}
\sup_{f\in\mathcal{M}_{g}}\text{Pr}_{f}\left(P_{g}\left(\mathbf{X}_{n}\right)\le\alpha\right)\le\alpha\text{,}\label{eq: P-value def}
\end{equation}
for any $\alpha\in\left(0,1\right)$. Here, $\text{Pr}_{f}$ is the
probability measure corresponding to the PDF $f$. In \citet{Wasserman:2020aa},
the following simple sequential testing procedure (STP) is proposed for
determining the value of $g_{0}$ (for general nested models, not
necessarily mixtures): 
\begin{enumerate}
\item Choose some significance level $\alpha\in\left(0,1\right)$ and initialize
$\hat{g}=0$;
\item Set $\hat{g}=\hat{g}+1$;
\item Test the null hypothesis $\text{H}_{\hat{g}}$ using the $p\text{-value}$
$P_{\hat{g}}\left(\mathbf{X}_{n}\right)$;
\begin{enumerate}
\item If $P_{\hat{g}}\left(\mathbf{X}_{n}\right)\le\alpha$, then go to Step 2.
\item If $P_{\hat{g}}\left(\mathbf{X}_{n}\right)>\alpha$, then go to Step 4.
\end{enumerate}
\item Output the estimated number of components $\hat{g}_{n}=\hat{g}$.
\end{enumerate}
It was argued informally in \citet{Wasserman:2020aa} that, although
the procedure above involves a sequence of multiple tests, each with
local size $\alpha$, it still correctly controls the Type I
error in the sense that 
\begin{equation}
\text{Pr}_{f_{0}}\left(f_{0}\in\mathcal{M}_{\hat{g}_{n}-1}\right)\le\alpha\label{eq: wasserman statement}
\end{equation}
for any $f_{0}\in\mathcal{M}$. Here, we note that the complement
of the event $\left\{ f_{0}\in\mathcal{M}_{\hat{g}_{n}-1}\right\} $
is $\left\{ f_{0}\in\mathcal{M}\backslash\mathcal{M}_{\hat{g}_{n}-1}\right\} $
or equivalently $\left\{ g_{0}\ge\hat{g}_{n}\right\} $. Thus, from
(\ref{eq: wasserman statement}), we can make the confidence
statement that
\begin{equation}
\text{Pr}_{f_{0}}\left(g_{0}\ge\hat{g}_{n}\right)\ge1-\alpha\text{,}\label{eq: confidence statement}
\end{equation}
for any $f_{0}\in\mathcal{M}$.

In the present work, we shall provide a formal proof of result
(\ref{eq: wasserman statement}) using the closed testing principle
of \citet{Marcus:1976aa} (see also
\citealp[Sec. 3.3]{Dickhaus2014}). Using this result and the universal
inference framework of \citet{Wasserman:2020aa}, we construct a sequence
of tests for $\left(\text{H}_{g}\right)_{g=1}^{\infty}$ with $p\text{-values}$
satisfying (\ref{eq: P-value def}) and prove that each of the tests is consistent under some regularity conditions. We then demonstrate the performance
of our testing procedure for the problem of order selection for finite
mixtures of normal distributions, and verify the empirical manifestation
of the confidence result (\ref{eq: confidence statement}). Extensions of the STP are also considered, whereupon we construct a method that consistently estimates the order $g_{0}$, and consider the application of the STP to asymptotically valid tests.

We note that hypothesis testing for order selection in mixture models
is a well-studied area of research. Difficulties in applying testing
procedures to the order selection problem arise due to identifiability
and boundary issues of the null hypothesis parameter spaces (see,
e.g., \citealp{Quinn:1987aa}, and references therein regarding parametric
mixture models, and \citealp{Andrews2001}, more generally). Examples
of testing methods proposed to overcome the problem include the parametric
bootstrapping techniques of \citet{McLachlan1987}, 
\citet{Feng1996}, and \citet{Polymenis:1998aa}, {whereupon bootstrapped distributions of test statistics are used to approximate finite sample distributions, in the absence of asymptotic results}. Another approach is the penalization
techniques of \citet{Chen1998}, \citet{Li:2010aa}, and \citet{Chen:2012aa}, {where asymptotically well-behaved penalized likelihood ratio statistics are proposed, with limiting distributions that are computable or simulatable. It is noteworthy that the bootstrap approaches provide only an approximate finite sample distribution of test statistics and thus the tests are not guaranteed to have the correct size. The penalization approach, on the other hand, provides asymptotic tests of the correct size, although the construction of the penalization of the test statistic must be specialized to every individual testing problem and is only suitable for parametric families of densities $\mathcal{K}$ that are characterized by a low-dimensional parameter.} 

In fact, the sequential procedure described
above was also considered for order selection in the mixture model
context by \citet{Windham:1992aa} and \citet{Polymenis:1998aa},
although no establishment of the properties of the approach was provided.
The possibility of constructing intervals of form (\ref{eq: confidence statement})
via bounding of discrete functionals of the underlying probability
measure is discussed in \citet{Donoho:1988aa}, although no implementation
is suggested. Citing observations made by \citet{Donoho:1988aa} and
\citet{Cutler:1994aa}, it is suggested in \citet[Sec. 6.1]{McLachlan2000}
that intervals of form (\ref{eq: confidence statement}) are sensible
in practice, because reasonable functionals that characterize properties
of $f_{0}$, such as for the number of components $g_{0}$, can be
lower bounded with high probability from data, but often cannot be upper bounded.

As previously mentioned, we plan to prove that (\ref{eq: wasserman statement})
holds by demonstrating that the sequential test is a closed testing
procedure. However, we note that the procedure may also be considered
under the sequential rejection principle of \citet{Goeman:2010aa},
and if $\mathcal{M}=\bigcup_{g=1}^{G}\mathcal{M}_{g}$ for some fixed
$G\in\mathbb{N}$, then we may also consider the procedure as a fixed
sequence procedure, as considered by \citet{Maurer:1995aa}. Another
perspective regarding the sequential test is via the general procedures
of \citet{Bauer:1996aa}, {who consider the construction of confidence intervals using sequences of tests for nested and order sets of hypotheses}. We also remark that the use of multiple
testing procedures for model selection is well studied in the literature,
as exemplified by the works of \citet{Finner:1996aa}
and \citet{Hansen:2011aa}, {who both consider the application of hypothesis testing schemes to generate confidence sets over model spaces.}

For completeness, we note that apart from hypothesis testing, numerous
solutions to the order selection problem for finite mixture models
have been suggested. These related works include the use of information
criteria, {such as the Akaike information crtierion (AIC), Bayesian information criterion (BIC), and variants of such techniques} \citep{Leroux1992,Biernacki2000,Keribin2000},
and parameter regularization, {such as via the Lasso and elastic net, and penalization approaches} \citep{Chen2009a,Xu:2015aa,Yin:2019aa},
among other techniques. 

{We note that the aforementioned order selection techniques are all, in a sense, point estimation procedures that each serve the purpose of consistently estimating the number of components of the DGP mixture model, in the sense that the estimate is close to the true number of components, for sufficiently large $n$. Our approach does not output a consistent estimator, but instead produces fixed-probability confidence set $\left\{ g_0\ge \hat{g}_{n}\right\}$, and should thus be viewed as an interval estimator. Although the lower-bound of the interval  $\hat{g}_n$ can be an accurate estimator of the true number of components, it should not be considered as a competitor to proper point estimators and instead should be viewed as complementary to point estimation approaches.} We finally note that outside of the multiple testing
framework, the problem of model selection with confidence has also
been addressed in the articles of \citet{Ferrari:2015aa} and \citet{Zheng:2019aa}.

The remainder of manuscript proceeds as follows. In Section 2, we
recall the closed testing principle and use it to prove the inequality
(\ref{eq: wasserman statement}). In Section 3, we use the universal
inference framework of \citet{Wasserman:2020aa} to construct a class
of likelihood ratio-based tests for the hypotheses $\left(\text{H}_{g}\right)_{g=1}^{\infty}$.
In the context of normal mixture models, numerical simulations {and real data examples} are
used to assess the performance of the sequential procedure using the
constructed tests in Section 4. Extensions to the STP are discussed in Section 5. Finally, conclusions are provided
in Section 6 {and technical proofs are provided in the Appendix}. 

\section{Confidence via the closed testing principle\label{sec:Confidence-via-the}}

Let $\mathbb{H}=\left\{ \text{H}_{g}:g\in\mathbb{G}\right\} $ be
a set of hypotheses that are indexed by some (possibly infinite) set
$\mathbb{G}$, where each hypothesis $\text{H}_{g}$ corresponds to
the statement $\left\{ \bm{\theta}\in\mathbb{T}_{g}\right\} $ regarding
the parameter of interest $\bm{\theta}\in\mathbb{T}$, where $\mathbb{T}_{g}\subset\mathbb{T}$.
We say that $\mathbb{H}$ is a $\cap\text{-closed}$ system if for
each $\mathbb{I}\subseteq\mathbb{G}$, either $\bigcap_{g\in\mathbb{I}}\mathbb{T}_{g}=\emptyset$
or $\bigcap_{g\in\mathbb{I}}\mathbb{T}_{g}\in\left\{ \mathbb{T}_{g}:g\in\mathbb{G}\right\} $.
That is, for every set $\mathbb{I}$ of indices that yields a non-empty statement $\left\{ \bm{\theta}\in\bigcap_{g\in\mathbb{I}}\mathbb{T}_{g}\right\} $, there exists a hypothesis $\text{H}_{g}\in\mathbb{H}$, such that $g\in\mathbb{T}_{g}$.

Recalling the notation from Section 1, we say that $\text{H}_{g}$
is rejected if $R_{g}\left(\mathbf{X}_{n}\right)=\mathbf{1}\left\{ P_{g}\left(\mathbf{X}_{n}\right)\le\alpha\right\} $
is equal to 1, and we say that $\text{H}_{g}$ is not rejected, otherwise.
Here, $\mathbf{1}\left\{ \cdot\right\} $ is the indicator function.
We further say that the familywise error rate (FWER) of a set rejections
$\left\{ R_{g}\left(\mathbf{X}_{n}\right)\right\} _{g\in\mathbb{G}}$ is strongly controlled
at level $\alpha\in\left(0,1\right)$ if for all $\bm{\theta}\in\mathbb{T}$,
\[
\text{Pr}_{\bm{\theta}}\left(\bigcup_{g\in\mathbb{G}_{0}\left(\bm{\theta}\right)}\left\{ R_{g}\left(\mathbf{X}_{n}\right)=1\right\} \right)\le\alpha\text{,}
\]
where $\text{Pr}_{\bm{\theta}}$ denotes the probability measure corresponding
to parameter value $\bm{\theta}$, and $\mathbb{G}_{0}\left(\bm{\theta}\right)\subset\mathbb{G}$
is the set of indices with corresponding hypotheses that  are
true under $\text{Pr}_{\bm{\theta}}$.

We note that the statement $\left\{ \bigcup_{g\in\mathbb{G}_{0}\left(\bm{\theta}\right)}\left\{ R_{g}\left(\mathbf{X}_{n}\right)=1\right\} \right\} $
reads as: at least one true hypothesis has been rejected. The complement
of the statement is therefore that no true hypotheses have been rejected
and hence the strong control of the FWER implies that the true parameter
value lies in the complement of union of the rejected subsets with
probability $1-\alpha$. That is, for all $\bm{\theta}\in\mathbb{T}$,
\[
\text{Pr}_{\bm{\theta}}\left(\bm{\theta}\in\bigcap_{g\in\mathbb{G}_{1}\left(\bm{X}_{n}\right)}\mathbb{T}_{g}^{\complement}\right)\ge1-\alpha\text{,}
\]
where $\left(\cdot\right)^{\complement}$ is the set complement operation and $\mathbb{G}_{1}\left(\bm{X}_{n}\right)=\left\{ g\in\mathbb{G}:R_{g}\left(\bm{X}_{n}\right)=1\right\}$ is the set of rejected hypotheses.

Define the set of closed tests corresponding to $\mathbb{H}$ as the
rejection rules: $\left(\bar{R}_{g}\left(\mathbf{X}_{n}\right)\right)_{g\in\mathbb{G}}$, where
for each $g\in\mathbb{G}$, 
\begin{equation}
\bar{R}_{g}\left(\mathbf{X}_{n}\right)=\min_{\left\{ j:\mathbb{T}_{j}\subseteq\mathbb{T}_{g}\right\} }R_{j}\left(\mathbf{X}_{n}\right)\text{.}\label{eq: closed tests}
\end{equation}
 all hypotheses $\text{H}_j, j \leq g$, are rejected, otherwise $\bar{R}_{g}\left(\mathbf{X}_{n}\right) = 0$. Then, we have the following result regarding the closed testing principle
(cf. \citealp[Thm. 3.4]{Dickhaus2014}).
\begin{thm}
\label{thm: closed tests}For an $\cap\text{-closed}$ system of hypotheses
$\mathbb{H}$ with corresponding $\alpha$ level local tests $\left(R_{g}\left(\mathbf{X}_{n}\right)\right)_{g\in\mathbb{G}}$,
the closed testing procedure defined by $\left(\bar{R}_{g}\left(\mathbf{X}_{n}\right)\right)_{g\in\mathbb{G}}$
strongly controls the FWER at level $\alpha$ in the sense that 
\[
\mathrm{Pr}_{\bm{\theta}}\left(\bigcup_{g\in\mathbb{G}_{0}\left(\bm{\theta}\right)}\left\{ \bar{R}_{g}\left(\mathbf{X}_{n}\right)=1\right\} \right)\le\alpha\text{,}
\]
for each $\bm{\theta}\in\mathbb{T}$.
\end{thm}
We now demonstrate that the sequential procedure constitutes a set
of closed tests of the form (\ref{eq: closed tests}) and thus permits
the conclusion of Theorem \ref{thm: closed tests}, which in turn
implies (\ref{eq: wasserman statement}) and thus (\ref{eq: confidence statement}). {That is, we show that the sequence of hypotheses $\left(\text{H}_{g}\right)_{g=1}^{\infty}$ corresponds to a $\cap$-closed system, where each $\text{H}_{g}$ is defined by $f_{0}\in\mathcal{M}_{g}$, and that the STP corresponds to a sequence of tests of form (\ref{eq: closed tests}).}
\begin{thm}
\label{thm: correctness of procedure}
The hypotheses $\left(\mathrm{H}_{g}\right)_{g=1}^{\infty}$ and the
STP from Section \ref{sec:Introduction}
constitute a $\cap\text{-closed}$ system and a closed testing procedure,
respectively, when testing using $p\text{-values}$ $\left(P_{g}\left(\mathbf{X}_{n}\right)\right)_{g=1}^{\infty}$,
satisfying (\ref{eq: P-value def}). The sequential test therefore
permit conclusions (\ref{eq: wasserman statement}) and (\ref{eq: confidence statement}).
\end{thm}
\begin{proof}
The proof of this result appears in the Appendix.
\end{proof}

{Thus, under the assumption that the data $\mathbf{X}_{n}$ arises from a DGP with density function $f_{0}$, corresponding to a $g_{0}$ component mixture model, the STP outputs a point estimator $\hat{g}_{n}$, where the event $\left\{ g_{0}\ge\hat{g}_{n}\right\}$  occurs with probability at least $1-\alpha$.}

\section{Test of order via universal
inference\label{sec:Local-order-testing}}

Let $\mathbf{X}_{n}$ be split into two subsequences of lengths $n_{1}$
and $n_{2}$, where $\mathbf{X}_{n}^{1}=\left(\bm{X}_{i}\right)_{i=1}^{n_{1}}$
and $\mathbf{X}_{n}^{2}=\left(\bm{X}_{i}\right)_{i=n_{1}+1}^{n}$,
and $n_{1}+n_{2}=n$. Assume that $\bm{X}$ has DGP characterized
by the PDF $f_{0}$ and for each $g\in\mathbb{N}$, let $\hat{f}_{g}^{1}\in\bar{\mathcal{M}}_{g}$
and $\hat{f}_{g}^{2}\in\bar{\mathcal{M}}_{g}$ be estimators of $f_{0}$
(not necessarily maximum likelihood estimators), based on $\mathbf{X}_{n}^{1}$
and $\mathbf{X}_{n}^{2}$, respectively, where $\bar{\mathcal{M}}_{g}\subseteq\mathcal{M}$ {is a class that characterizes an alternative to the null hypothesis that $f_{0}\in\mathcal{M}_{g}$, with $\mathcal{M}_{g}\subset\bar{\mathcal{M}}_{g}$}.

For notational convenience,
for each $k\in\left\{ 1,2\right\} $, we reindex the elements of $\mathbf{X}_n^k$ by inclusion of a superscript $k$, so that  $\mathbf{X}_{n}^{k}=\left(\bm{X}_{i}^{k}\right)_{i=1}^{n_{k}}$, and let
\[
L_{f}\left(\mathbf{X}_{n}^{k}\right)=\prod_{i=1}^{n_{k}}f\left(\bm{X}_{i}^{k}\right)\text{,}
\]
be the likelihood function corresponding to subsample $\mathbf{X}_{n}^{k}$,
evaluated under PDF $f$. We wish to test the null hypothesis $\text{H}_{g}$:
$f_{0}\in\mathcal{M}_{g}$ against the alternative $\bar{\text{H}}_{g}$:
$f_{0}\in\bar{\mathcal{M}}_{g}$, using the Split test statistics
\[
V_{g}^{k}\left(\mathbf{X}_{n}\right)=\frac{L_{\hat{f}_{g}^{3-k}}\left(\mathbf{X}_{n}^{k}\right)}{L_{\tilde{f}_{g}^{k}}\left(\mathbf{X}_{n}^{k}\right)}\text{,}
\]
for $k\in\left\{ 1,2\right\} $, and the Swapped test statistic
\[
\bar{V}_{g}\left(\mathbf{X}_{n}\right)=\frac{1}{2}\left\{ V_{g}^{1}\left(\mathbf{X}_{n}\right)+V_{g}^{2}\left(\mathbf{X}_{n}\right)\right\} \text{,}
\]
{as introduced in \cite{Wasserman:2020aa}}. Here, the denominator estimator $\tilde{f}_{g}^{k}$ is the maximum likelihood estimator of $f_{0}$,
based on $\mathbf{X}_{n}^{k}$ under the null hypothesis $\text{H}_{g}$,
in the sense that

\[
\tilde{f}_{g}^{k}\in\left\{ \tilde{f}\in\mathcal{M}_{g}:L_{\tilde{f}}\left(\mathbf{X}_{n}^{k}\right)=\max_{f\in\mathcal{M}_{g}}L_{f}\left(\mathbf{X}_{n}^{k}\right)\right\} \text{.}
\]

We define the $p\text{-values}$ for the Split and Swapped test statistics
as $P_{g}^{k}\left(\mathbf{X}_{n}\right)=\min\{1/V_{g}^{k}\left(\mathbf{X}_{n}\right),1\}$
and $\bar{P}_{g}\left(\mathbf{X}_{n}\right)=\min\{1/\bar{V}_{g}\left(\mathbf{X}_{n}\right),1\}$,
respectively. The adaptation of \citet[Thm. 3]{Wasserman:2020aa}
demonstrates that the two tests have correct size for any sample size
$n$ (i.e., $P_{g}^{k}\left(\mathbf{X}_{n}\right)$ and $\bar{P}_{g}\left(\mathbf{X}_{n}\right)$
satisfy condition (\ref{eq: P-value def}), for any $n$).
\begin{thm}
\label{thm: type 1 control}For any $n\in\mathbb{N}$ and $\alpha\in\left(0,1\right)$,
\[
\sup_{f\in\mathcal{M}_{g}}\mathrm{Pr}_{f}\left(P_{g}^{k}\left(\mathbf{X}_{n}\right)\le\alpha\right)\le\alpha
\]
and 
\[
\sup_{f\in\mathcal{M}_{g}}\mathrm{Pr}_{f}\left(\bar{P}_{g}\left(\mathbf{X}_{n}\right)\le\alpha\right)\le\alpha\text{.}
\]
\end{thm}

{Theorem 3 implies that for each $g\in\mathbb{N}$ and $k\in\left\{ 1,2\right\}$, and for any sample size $n\in\mathbb{N}$, if $f_{0}\in\mathcal{M}_{g}$ is the DGP of $\mathbf{X}_{n}$, then events $\left\{ P_{g}^{k}\left(\mathbf{X}_{n}\right)\le\alpha\right\}$  and $\left\{ \bar{P}_{g}\left(\mathbf{X}_{n}\right)\le\alpha\right\}$, corresponding to a rejection of the null hypothesis $\text{H}_{g}$, occur with probability no greater than $\alpha$, as required for a test of size $\alpha$.}

It is suggested by \citet{Windham:1992aa}, \citet{Polymenis:1998aa},
and \citet{Wasserman:2020aa} that the alternative hypothesis for
each $\text{H}_{g}$ should be that $f_{0}\in\bar{\mathcal{M}}_{g}=\mathcal{M}_{g+1}$.
However, since we are only looking to reject $\text{H}_{g}$, rather
than making conclusions regarding the alternative, we can take $\bar{\mathcal{M}}_{g}$
to be a richer class of PDFs that is still feasible to estimate. Thus,
in the sequel, we shall consider the possibility that $\bar{\mathcal{M}}_{g}=\mathcal{M}_{g+l_{g}}$
for some $l_{g}\in\mathbb{N}$, for each $g\in\mathbb{N}$.
Typically, we can let $l_{g}=l$ for all $g$, but we anticipate that
there may be circumstances where one may wish for $l_{g}$ to vary.

\subsection{Consistency of order tests}

Although Theorem \ref{thm: type 1 control} guarantees the control
of the Type I error for each local test of $\text{H}_{g}$, it makes
no statement regarding the power of the tests. For tests against alternatives
of the form: $\bar{\mathcal{M}}_{g}=\mathcal{M}_{g+l_{g}}$, we shall
consider the issue of power from an asymptotic perspective in the
parametric context. That is, we suppose that 
\begin{align}
    \mathcal{K}\left(\mathbb{X}\right)=\left\{ f\left(\bm{x}\right)=f\left(\bm{x};\bm{\theta}\right):\bm{\theta}\in\mathbb{T}\right\} \text{,} \label{eq: K Parametric}
\end{align}
where $\mathbb{T}\subseteq\mathbb{R}^{p}$ for some $p\in\mathbb{N}$,
and thus
\begin{align}
\mathcal{M}_{g}\left(\mathbb{X}\right)=\left\{ f\left(\bm{x};\bm{\vartheta}^{\left(g\right)}\right):f\left(\bm{x};\bm{\vartheta}^{\left(g\right)}\right)=\sum_{z=1}^{g}\pi_{z}f\left(\bm{x};\bm{\theta}_{z}\right);\pi_{z}\ge0,\sum_{z=1}^{g}\pi_{z}=1,\bm{\theta}_{g}\in\mathbb{T},z\in\left[g\right]\right\} \text{.} \label{eq: Mg Parametric}
\end{align}

We put the pairs $\left(\left(\pi_{z},\bm{\theta}_{z}\right)\right)_{z=1}^{g}$
in the vector $\bm{\vartheta}^{\left(g\right)}\in\left(\left[0,1\right]\times\mathbb{T}\right)^{g}=\mathbb{T}_{g}$.
Here, we further replace $\hat{f}_{g}^{2}$ and $\tilde{f}_{g}^{1}$
by $f\left(\cdot;\hat{\bm{\vartheta}}_{n}^{\left(g+l_{g}\right)}\right)$
and $f\left(\cdot;\tilde{\bm{\vartheta}}_{n}^{\left(g\right)}\right)$,
respectively, where $\hat{\bm{\vartheta}}_{n}^{\left(g+l_{g}\right)}$
is a function of $\mathbf{X}_{n}^{2}$ and $\tilde{\bm{\vartheta}}_{n}^{\left(g\right)}$
is a function of $\mathbf{X}_{n}^{1}$. Further, since $f\left(\cdot;\tilde{\bm{\vartheta}}_{n}^{\left(g\right)}\right)$
is the maximum likelihood estimator of $f_{0}\in\mathcal{M}_{g}$,
we also write
\begin{equation}
\tilde{\bm{\vartheta}}_{n}^{\left(g\right)}\in\left\{ \tilde{\bm{\vartheta}}^{\left(g\right)}\in\mathbb{T}_{g}:\prod_{i=1}^{n_{1}}f\left(\bm{X}_{i}^{1};\tilde{\bm{\vartheta}}^{\left(g\right)}\right)=\max_{\bm{\vartheta}^{\left(g\right)}\in\mathbb{T}_{g}}\prod_{i=1}^{n_{1}}f\left(\bm{X}_{i}^{1};\bm{\vartheta}^{\left(g\right)}\right)\right\} \text{.}\label{eq: mle}
\end{equation}

Following \citet[Def. 23.1]{DasGupta2008}, we say that a sequence of tests
$\left(R_{g}\left(\mathbf{X}_{n}\right)\right)_{n=1}^{\infty}$ for $\text{H}_{g}$ is consistent if under the true DGP,
characterized by $f_{0}\notin\mathcal{M}_{g}$, it is true that $\text{Pr}_{f_{0}}\left(R_{g}\left(\mathbf{X}_{n}\right)=1\right)\rightarrow1$,
as $n\rightarrow\infty$. Let $\left\Vert \cdot\right\Vert $ denote
the Euclidean norm and define the Kullback--Leibler divergence between
two PDFs on $\mathbb{X}$: $f_{1}$ and $f_{2}$, as
\[
\text{D}\left(f_{1},f_{2}\right)=\int_{\mathbb{X}}f_{1}\left(\bm{x}\right)\log\frac{f_{1}\left(\bm{x}\right)}{f_{2}\left(\bm{x}\right)}\text{d}\bm{x}\text{.}
\]
Further, say that a class of parametric mixture models $\mathcal{M}_{g}$
is identifiable if
\[
\sum_{z=1}^{g}\pi_{z}f\left(\bm{x};\bm{\theta}_{z}\right)=\sum_{z=1}^{g}\pi_{z}^{\prime}f\left(\bm{x};\bm{\theta}_{z}^{\prime}\right)
\]
if and only if $\sum_{z=1}^{g}\pi_{z}\mathbf{1}\left(\bm{\theta}=\bm{\theta}_{z}\right)=\sum_{z=1}^{g}\pi_{z}^{\prime}\mathbf{1}\left(\bm{\theta}=\bm{\theta}_{z}^{\prime}\right)$,
where $\text{\textbf{1}}\left(\cdot\right)$ is the usual indicator
function. For $R_{g}\left(\mathbf{X}_{n}\right)=\mathbf{1}\left(P_{g}^{1}\left(\mathbf{X}_{n}\right)<\alpha\right)$,
where $P_{g}^{1}\left(\mathbf{X}_{n}\right)$ is obtained from testing
$\text{H}_{g}$ against the alternative $\bar{\mathcal{M}}_{g}=\mathcal{M}_{g+l_{g}}$,
we obtain the following result. The equivalent result regarding $\bar{P}_{g}\left(\bm{X}_{n}\right)$ can be established analogously.
\begin{thm}
\label{thm: individual test consistency}
Make the following assumptions:
\begin{description}
\item [{(A1)}] for each $g\in\mathbb{N}$, the class
$\mathcal{M}_{g}$ is identifiable;
\item [{(A2)}] the PDF $f\left(\bm{x};\bm{\theta}\right)>0$ is everywhere
positive and continuous for all $\left(\bm{x},\bm{\theta}\right)\in\mathbb{X}\times\mathbb{T}$,
where $\mathbb{X}$ and $\mathbb{T}$ are Euclidean spaces and $\mathbb{T}$
is compact;
\item [{(A3)}] for all $\bm{x}\in\mathbb{X}$ and $\bm{\theta}_{1},\bm{\theta}_{2}\in\mathbb{T}$,
$\left|\log f\left(\bm{x};\bm{\theta}_{1}\right)\right|\le M_{1}\left(\bm{x}\right)$
and
\[
\left|\log f\left(\bm{x};\bm{\theta}_{1}\right)-\log f\left(\bm{x};\bm{\theta}_{2}\right)\right|\le M_{2}\left(\bm{x}\right)\left\Vert \bm{\theta}_{1}-\bm{\theta}_{2}\right\Vert ,
\]
where $\mathrm{E}_{f_{0}}M_{1}\left(\bm{X}\right)<\infty$ and $\mathrm{E}_{f_{0}}M_{2}\left(\bm{X}\right)<\infty$;
\item [{(A4)}] the estimator $\hat{\bm{\vartheta}}_{n}^{\left(g+l_{g}\right)}\rightarrow\bm{\vartheta}_{0}^{\left(g+l_{g}\right)}$,
in probability, as $n_{2}\rightarrow\infty$, where 
\[
\bm{\vartheta}_{0}^{\left(g+l_{g}\right)}\in\left\{ \hat{\bm{\vartheta}}^{\left(g+l_{g}\right)}\in\mathbb{T}_{g+l_{g}}:\mathrm{E}_{f_{0}}\log f\left(\bm{X};\hat{\bm{\vartheta}}^{\left(g+l_{g}\right)}\right)=\max_{\bm{\vartheta}^{\left(g+l_{g}\right)}\in\mathbb{T}_{g+l_{g}}}\mathrm{E}_{f_{0}}\log f\left(\bm{X};\bm{\vartheta}^{\left(g+l_{g}\right)}\right)\right\} \text{.}
\]
\end{description}
Under Assumptions (A1)--(A4), if $f_{0}\in\mathcal{M}\backslash\mathcal{M}_{g}$,
and $n_{1},n_{2}\rightarrow\infty$, then $R_{g}\left(\mathbf{X}_{n}\right)=\mathbf{1}\left(P_{g}^{1}\left(\mathbf{X}_{n}\right)<\alpha\right)$
is a consistent test for $\text{H}_{g}$.
\end{thm}
\begin{proof}
The proof of this result appears in the Appendix.
\end{proof}

{
Assumption A1 ensures that the elements of (\ref{eq: Mg Parametric}) (i.e., the $g$ component mixtures of densities of class (\ref{eq: K Parametric})) are distinct (as noted in \citealt[Sec. 3.1.1]{Titterington1985}), and A2 implies that the log-likelihood cannot take infinitely negative values and that it is continuous for any $\bm{x}$ and $\bm{\theta}$, where the compactness of $\mathbb{T}$ ensures that $f\left(\bm{x};\bm{\theta}\right)$ is bounded for each fixed $\bm{x}\in\mathbb{X}$. Assumption A3 then implies that the expected log-likelihood $\text{E}_{f_{0}}\log f\left(\bm{X};\bm{\theta}\right)$ is bounded for each $\bm{\theta}$, and since $f\left(\bm{x};\bm{\theta}\right)$ is continuous, $\text{E}_{f_{0}}\log f\left(\bm{X};\bm{\theta}\right)$ is also continuous and thus has global optima within the compact set $\mathbb{T}$. Assumption A3 also implies that $\text{E}_{f_{0}}\log f\left(\bm{X};\bm{\theta}\right)$ is Lipschitz continuous, with respect to $\bm{\theta}\in\mathbb{T}$ equipped with the Euclidean norm, and A4 implies that $\hat{\bm{\vartheta}}_{n}^{\left(g+l_{g}\right)}$, characterizing $\hat{f}_{g}^{2}$, behaves asymptotically (with respect to convergence in probability) like a parametric maximum likelihood estimator, under the potentially misspecified supposition that  $f_{0}\in\mathcal{M}_{g+l_{g}}$.}

{
Assumptions A1--A3 are required for the application of \citet[Lem. 1]{Leroux1992}, with A2 and A3 also required for establishing the consistency of the estimators $\tilde{\bm{\vartheta}}_{n}^{\left(g\right)}$. The Lipschitz condition of A3 and A4 are further required to show that the logarithm of the split test statistic $V_{g}^{1}\left(\mathbf{X}_{n}\right)$ is a consistent estimator of an difference in divergence expression required in the proof.
}

{Theorem \ref{thm: individual test consistency} states that for each $g<g_0$ and for any significance level $\alpha\in(0,1)$, the rejection probability of the test of $\text{H}_g$, based on $P_{g}^{1}\left(\mathbf{X}_{n}\right)$, converges to 1, as $n$ gets large.} We note that Assumptions A1--A4 are verifiable for typical models of interest. For example, when $\mathcal{M}_{g}$ is the class of $g$ component normal mixture models (see Section 4), A1 is verified due to \cite{Yakowitz1968}, A2 is satisfied by the usual compact restrictions on the parameter space (see, e.g., \citealp[Sec. B.6.2]{ritter2014robust}), and A3 is satisfied under A2. 

\section{Normal mixture models}

We apply the STP with the Split and Swapped tests from Section \ref{sec:Local-order-testing}
to the classic problem of order selection for normal mixture models,
whereby $\mathbb{X}=\mathbb{R}^{d}$ for some $d\in\mathbb{N}$ and
\[
\mathcal{K}\left(\mathbb{X}\right)=\mathcal{K}\left(\mathbb{R}^{d}\right)=\left\{ f\left(\bm{x}\right)=\phi\left(\bm{x};\bm{\mu},\bm{\Sigma}\right):\phi\left(\bm{x};\bm{\mu},\bm{\Sigma}\right)=\left|2\pi\bm{\Sigma}\right|^{-1/2}\exp\left[-\frac{1}{2}\left(\bm{x}-\bm{\mu}\right)^{\top}\bm{\Sigma}^{-1}\left(\bm{x}-\bm{\mu}\right)\right]\right\} \text{,}
\]
where $\bm{\mu}\in\mathbb{R}^{d}$ and $\bm{\Sigma}\in\mathbb{R}^{d\times d}$
is symmetric positive definite.

\begin{figure}
\centering\includegraphics[width=15cm]{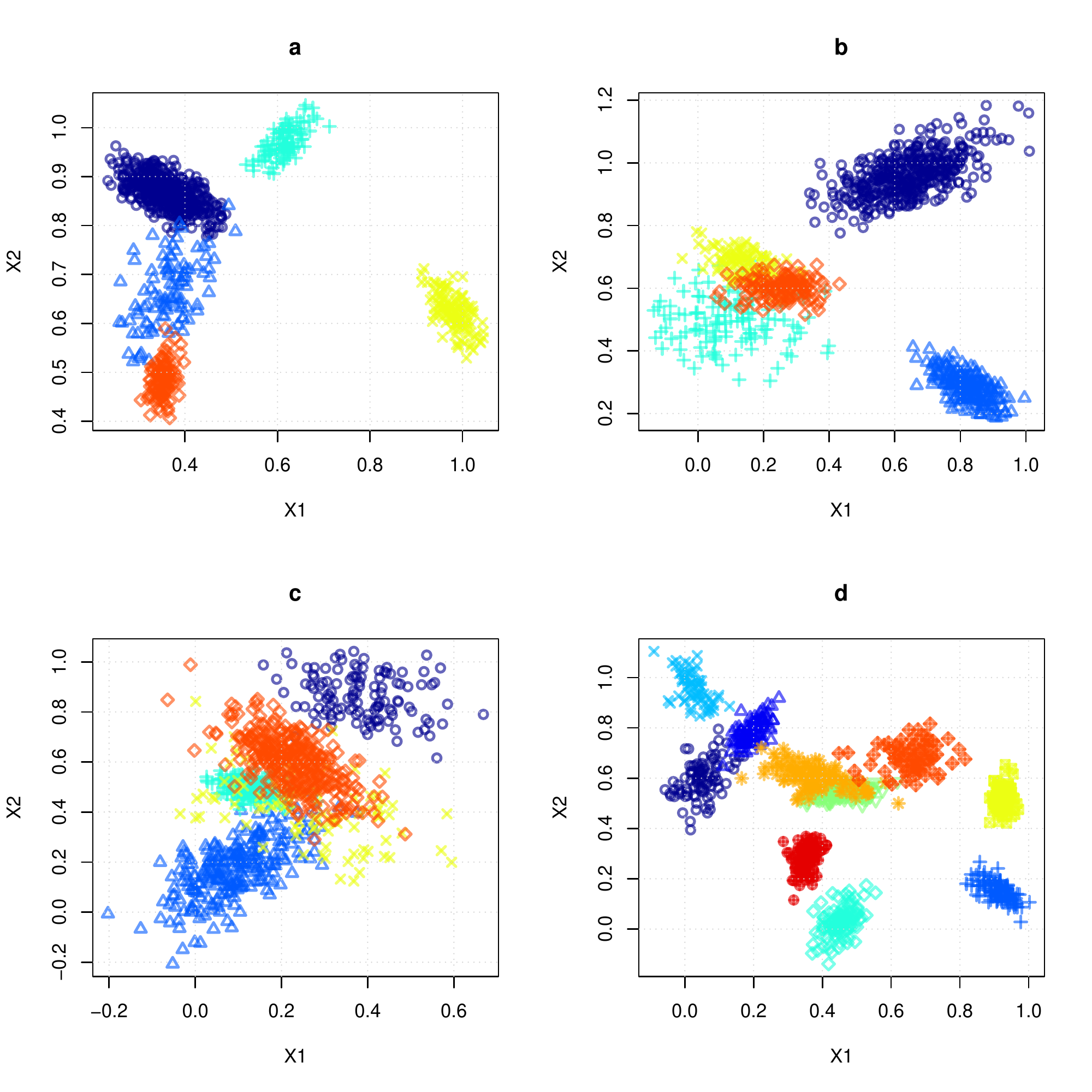}

\caption{\label{fig: Data Example} Example data sets of $n_{1}=1000$ random
observations from a $d=2$ dimensional $g_{0}$ component normal mixture
model, with parameters determined via parameter $\bar{\omega}$. Here,
the pairs $\left(g_{0},\bar{\omega}\right)$ visualized in subplots a, b, c,
and d are $\left(5,0.01\right)$, $\left(5,0.05\right)$, $\left(5,0.1\right)$,
and $\left(10,0.01\right)$, respectively.}
\end{figure}

To assess the performance of the STP, we conduct a thorough simulation
study, within the $\mathsf{R}$ programming environment \citep{R-Core-Team:2020aa}.
For each $d\in\left\{ 2,4\right\} $, we generate data sets $\mathbf{X}_{n}$,
with $n_{1}=n_{2}\in\left\{{300,500,}1000,2000,5000,10000\right\} $ observations
(recall that $n=n_{1}+n_{2}$), where each $\bm{X}_{i}\in\mathbb{R}^{d}$,
from a multivariate normal mixture model in $\mathcal{M}_{g_{0}}\left(\mathbb{R}^{d}\right)$
for $g_{0}\in\left\{ 5,10\right\} $, with parameter elements $\left(\pi_{z},\bm{\mu}_{z},\bm{\Sigma}_{z}\right)_{z=1}^{g_{0}}$
of $\mathcal{M}_{g_{0}}\left(\mathbb{R}^{d}\right)$ generated using
the $\mathsf{MixSim}$ package \citep{Melnykov:2012aa}, using the
setting $\bar{\omega}\in\left\{ 0.01,0.05,0.1\right\} $ and $\min_{z\in\left[g_{0}\right]}\pi_{z}\ge\left(2g_{0}\right)^{-1}$.
Here, the $\bar{\omega}$ parameter is described in \citet{Melnykov:2012aa}, and controls the level of overlap between
the normal components of the mixture model. Four examples of data
sets generated using various combinations of simulation parameters
$\left(g_{0},\bar{\omega}\right)$, with $d=2$ and $n_{1}=1000$,
are provided in Figure \ref{fig: Data Example}.

For each set of simulation parameters $\left(g_{0},\bar{\omega},d,n_{1}\right)$,
we simulate $r=100$ replicate data sets, whereupon we apply the STP
at the $\alpha=0.05$ level, using the Split and Swapped test $p$-values
of the forms $P_{g}^{1}$ and $\bar{P}_{g}$, for each of the $r$
data sets. To compute the maximum likelihood estimators $\tilde{f}_{g}^{k}=f\left(\cdot;\tilde{\bm{\vartheta}}_{n}^{\left(g\right)}\right)$,
under the null hypotheses that $f_{0}\in\mathcal{M}_{g}$, we use
the $\texttt{gmm\_full}$ function from the $\textsf{Armadillo}$
$\textsf{C++}$ library, implemented in $\textsf{R}$ using the $\textsf{RcppArmadillo}$
package \citep{Eddelbuettel:2014vf}. We also use the maximum likelihood
estimator as $\hat{f}_{g}^{3-k}=f\left(\cdot;\hat{\bm{\vartheta}}_{n}^{\left(g+l_{g}\right)}\right)$,
under the alternative hypotheses $f_{0}\in\bar{\mathcal{M}}_{g}=\mathcal{M}_{g+l_{g}}$,
where we set $l_{g}=l\in\left\{ 1,2\right\} $, for all $g\in\mathbb{N}$.
From each of the $r$ STP results, we compute the coverage proportion
(CovProp; proportion of $r$ for which $g_{0}\ge\hat{g}_{n}$), the
mean estimated number of components (MeanComp; the average of $\hat{g}_{n}$
over the $r$ repetitions), and the proportion of times that the estimated
number of components corresponded with the $g_{0}$ (CorrProp; the
proportion of times the event $\hat{g}_{n}=g_{0}$ occurs out of the
$r$ repetitions).

{
It is worth recalling that our estimators $\hat{g}_{n}$ are not in
fact point estimators of $g_{0}$, but are actually lower bounds of the STP 
interval estimators $\left\{ g_{0}\ge\hat{g}_{n}\right\} $ and are
thus only expected to be close $g_{0}$, with $\hat{g}_{n}=g_{0}$ indicating
that the interval is efficient. We can complement the output of the
interval estimator with a point estimator of $g_{0}$, such as via
the AIC or BIC procedures, whereupon for data set $\mathbf{X}_{n}$,
and for each $g\in\mathbb{N}$, the AIC or BIC values:
\[
\text{AIC}_{g}=\frac{2}{n}\text{dim}_{g}-\frac{2}{n}\sum_{i=1}^{n}\log f\left(\bm{X}_{i};\bm{\vartheta}_{n}^{\left(g\right)}\right)\text{, and}
\]
\begin{align}
\text{BIC}_{g}=\frac{\log n}{n}\text{dim}_{g}-\frac{2}{n}\sum_{i=1}^{n}\log f\left(\bm{X}_{i};\bm{\vartheta}_{n}^{\left(g\right)}\right)\text{,} \label{eq: BIC def}
\end{align}
respectively, are computed. Here $\text{dim}_g$ is the dimensionality of $\bm{\vartheta}^{(g)}$ and $\bm{\vartheta}^{(g)}_n$ is a maximum likelihood estimator of form (\ref{eq: mle}), computed using $\boldsymbol{X}_n$ instead of $\boldsymbol{X}_n^1$. The AIC and BIC procedures then estimate
$g_{0}$ via $\arg\min_{g}\text{AIC}_{g}$ or $\arg\min_{g}\text{BIC}_{g}$,
respectively. To complement our results regarding $\left\{ g_{0}\ge\hat{g}_{n}\right\} $,
we also provide the MeanComp and CorrProp values for the AIC and BIC
procedures.
}
All of our \textsf{R} scripts are made available at \url{https://github.com/ex2o/oscfmm}.

\subsection{Simulation results}

For all scenario combinations $\left(g_{0},\bar{\omega},d,n_{1},l\right)$,
the CovProp was $100\%$ over the $r$ repetitions. This confirms
the conclusions of Theorems \ref{thm: correctness of procedure} and
\ref{thm: type 1 control}. This also implies that the tests are underpowered,
which is conforming to the observations from the simulations of \citet{Wasserman:2020aa}.
This result is unsurprising since the tests are constructed via a
Markov inequality argument, which makes no use of the topological
features of the sets $\mathcal{M}_{g}$ and $\bar{\mathcal{M}}_{g}$
that can be used to derive more specific results.

{We report the STP interval estimator, and AIC and BIC point estimator results} for all of the combinations
$\left(g_{0},\bar{\omega},d,n_{1},l\right)$, partitioned by $\left(g_{0},\bar{\omega}\right)$
in Tables \ref{tab: 5,0.01}--\ref{tab: 10,0.1}. Here, Tables \ref{tab: 5,0.01}--\ref{tab: 10,0.1}
contain results for pairs $\left(5,0.01\right)$, $\left(5,0.05\right)$,
$\left(5,0.1\right)$, $\left(10,0.01\right)$, $\left(10,0.05\right)$,
and $\left(10,0.1\right)$, respectively. In the $\left(5,0.01\right)$
case, we observe that both the Split and Swapped test-based STPs were
able to identify the generative value of $g$ in over $90\%$ of the
cases, except when {$n_{1}<1000$ and for the case} $\left(d,n_{1},l\right)=\left(4,1000,2\right)$.
There is some evidence that the Swapped test is more powerful than
Split test in all cases, as indicated by the higher values of MeanComp
and CorrProp. Furthermore, the $l=2$ alternative appears to be more
powerful than the $l=1$ alternative in all cases except when $\left(d,n_{1}\right)=\left(4,1000\right)$.

\begin{table}[ht]
\caption{\label{tab: 5,0.01}{MeanComp and CorrProp results for different values
of $\left(d,n_{1},l\right)$, when $\left(g_{0},\bar{\omega}\right)=\left(5,0.01\right)$.}}

\centering %
\begin{tabular}{|rrr|rrrr|rrrr|}
\hline 
 &  &  &  & MeanComp  &  &  &  & CorrProp &  & \tabularnewline
$d$  & $n_{1}$  & $l$  & Split  & Swapped  & AIC & BIC & Split  & Swapped & AIC & BIC\tabularnewline
\hline 
\hline 
2  & 300  & 1  & 4.25  & 4.52  & 4.99  & 4.49  & 0.50  & 0.65 & 0.49  & 0.50 \tabularnewline
 &  & 2  & 4.53  & 4.63  &  &  & 0.56  & 0.63 &  & \tabularnewline
 & 500  & 1  & 4.69  & 4.89  & 4.95  & 4.50  & 0.82  & 0.92 & 0.52  & 0.50 \tabularnewline
 &  & 2  & 4.81  & 4.93  &  &  & 0.81  & 0.93 &  & \tabularnewline
 & 1000  & 1  & 4.88  & 4.95  & 5.20  & 5.00  & 0.95  & 0.98 & 0.87  & 1.00 \tabularnewline
 &  & 2  & 4.92  & 4.96  &  &  & 0.92  & 0.96 &  & \tabularnewline
 & 2000  & 1  & 4.89  & 4.94  & 5.13  & 5.00  & 0.96  & 0.98 & 0.88  & 1.00 \tabularnewline
 &  & 2  & 5.00  & 5.00  &  &  & 1.00  & 1.00 &  & \tabularnewline
 & 5000  & 1  & 4.94  & 4.98  & 5.05  & 5.00  & 0.97  & 0.99 & 0.96  & 1.00 \tabularnewline
 &  & 2  & 5.00  & 5.00  &  &  & 1.00  & 1.00 &  & \tabularnewline
 & 10000  & 1  & 4.93  & 4.96  & 5.05  & 5.00  & 0.97  & 0.98 & 0.95  & 1.00 \tabularnewline
 &  & 2  & 4.99  & 4.99  &  &  & 0.99  & 0.99 &  & \tabularnewline
4  & 300  & 1  & 3.45  & 3.87  & 5.38  & 4.46  & 0.08  & 0.18 & 0.42  & 0.47 \tabularnewline
 &  & 2  & 4.03  & 4.13  &  &  & 0.11  & 0.16 &  & \tabularnewline
 & 500  & 1  & 4.31  & 4.47  & 5.22  & 4.50  & 0.43  & 0.53 & 0.40  & 0.49 \tabularnewline
 &  & 2  & 4.34  & 4.41  &  &  & 0.34  & 0.41 &  & \tabularnewline
 & 1000  & 1  & 4.87  & 4.97  & 5.25  & 5.00  & 0.92  & 0.97 & 0.79  & 1.00 \tabularnewline
 &  & 2  & 4.85  & 4.86  &  &  & 0.85  & 0.86 &  & \tabularnewline
 & 2000  & 1  & 4.94  & 4.98  & 5.16  & 5.00  & 0.97  & 0.99 & 0.88  & 1.00 \tabularnewline
 &  & 2  & 5.00  & 5.00  &  &  & 1.00  & 1.00 &  & \tabularnewline
 & 5000  & 1  & 4.98  & 5.00  & 5.03  & 5.00  & 0.99  & 1.00 & 0.98  & 1.00 \tabularnewline
 &  & 2  & 5.00  & 5.00  &  &  & 1.00  & 1.00 &  & \tabularnewline
 & 10000  & 1  & 4.98  & 4.98  & 5.07  & 5.00  & 0.99  & 0.99 & 0.93  & 1.00 \tabularnewline
 &  & 2  & 5.00  & 5.00  &  &  & 1.00  & 1.00 &  & \tabularnewline
\hline 
\end{tabular}
\end{table}

\begin{table}[ht]
\caption{{MeanComp and CorrProp results for different values of $\left(d,n_{1},l\right)$,
when $\left(g_{0},\bar{\omega}\right)=\left(5,0.05\right)$.}}

\centering %
\begin{tabular}{|rrr|rrrr|rrrr|}
\hline 
 &  &  &  & MeanComp  &  &  &  & CorrProp &  & \tabularnewline
$d$  & $n_{1}$  & $l$  & Split  & Swapped  & AIC & BIC & Split  & Swapped & AIC & BIC\tabularnewline
\hline 
\hline 
2  & 300  & 1  & 3.60  & 3.77  & 5.16  & 4.17  & 0.11  & 0.14 & 0.40  & 0.36 \tabularnewline
 &  & 2  & 3.87  & 4.02  &  &  & 0.09  & 0.13 &  & \tabularnewline
 & 500  & 1  & 3.89  & 4.09  & 5.08  & 4.30  & 0.26  & 0.35 & 0.47  & 0.41 \tabularnewline
 &  & 2  & 4.20  & 4.32  &  &  & 0.23  & 0.35 &  & \tabularnewline
 & 1000  & 1  & 4.41  & 4.52  & 5.13  & 4.92  & 0.57  & 0.62 & 0.89  & 0.92 \tabularnewline
 &  & 2  & 4.51  & 4.60  &  &  & 0.53  & 0.60 &  & \tabularnewline
 & 2000  & 1  & 4.82  & 4.92  & 5.10  & 4.96  & 0.88  & 0.94 & 0.91  & 0.96 \tabularnewline
 &  & 2  & 4.85  & 4.88  &  &  & 0.85  & 0.88 &  & \tabularnewline
 & 5000  & 1  & 4.96  & 4.96  & 5.13  & 5.00  & 0.98  & 0.98 & 0.89  & 1.00 \tabularnewline
 &  & 2  & 4.90  & 4.93  &  &  & 0.90  & 0.93 &  & \tabularnewline
 & 10000  & 1  & 4.92  & 4.92  & 5.02  & 5.00  & 0.96  & 0.96 & 0.99  & 1.00 \tabularnewline
 &  & 2  & 4.95  & 4.98  &  &  & 0.97  & 0.98 &  & \tabularnewline
4  & 300  & 1  & 2.73  & 2.83  & 5.63  & 3.89  & 0.00  & 0.00 & 0.40  & 0.25 \tabularnewline
 &  & 2  & 3.00  & 3.12  &  &  & 0.00  & 0.00 &  & \tabularnewline
 & 500  & 1  & 3.09  & 3.42  & 5.55  & 4.21  & 0.02  & 0.04 & 0.38  & 0.36 \tabularnewline
 &  & 2  & 3.72  & 3.84  &  &  & 0.04  & 0.03 &  & \tabularnewline
 & 1000  & 1  & 3.98  & 4.25  & 5.41  & 4.95  & 0.22  & 0.37 & 0.72  & 0.95 \tabularnewline
 &  & 2  & 4.20  & 4.27  &  &  & 0.21  & 0.27 &  & \tabularnewline
 & 2000  & 1  & 4.71  & 4.85  & 5.19  & 4.98  & 0.77  & 0.87 & 0.85  & 0.98 \tabularnewline
 &  & 2  & 4.73  & 4.79  &  &  & 0.73  & 0.79 &  & \tabularnewline
 & 5000  & 1  & 4.96  & 5.00  & 5.10  & 5.00  & 0.98  & 1.00 & 0.91  & 1.00 \tabularnewline
 &  & 2  & 4.96  & 4.98  &  &  & 0.96  & 0.98 &  & \tabularnewline
 & 10000  & 1  & 4.98  & 4.98  & 5.06  & 5.00  & 0.99  & 0.99 & 0.94  & 1.00 \tabularnewline
 &  & 2  & 4.99  & 5.00  &  &  & 0.99  & 1.00 &  & \tabularnewline
\hline 
\end{tabular}

\end{table}

\begin{table}[ht]
\caption{{MeanComp and CorrProp results for different values of $\left(d,n_{1},l\right)$,
when $\left(g_{0},\bar{\omega}\right)=\left(5,0.1\right)$.}}

\centering %
\begin{tabular}{|rrr|rrrr|rrrr|}
\hline 
 &  &  &  & MeanComp  &  &  &  & CorrProp &  & \tabularnewline
$d$  & $n_{1}$  & $l$  & Split  & Swapped  & AIC & BIC & Split  & Swapped & AIC & BIC\tabularnewline
\hline 
\hline 
2  & 300  & 1  & 2.85  & 3.16  & 5.08  & 3.63  & 0.00  & 0.02 & 0.43  & 0.17 \tabularnewline
 &  & 2  & 3.19  & 3.35  &  &  & 0.01  & 0.01 &  & \tabularnewline
 & 500  & 1  & 3.33  & 3.65  & 5.01  & 4.02  & 0.06  & 0.09 & 0.43  & 0.24 \tabularnewline
 &  & 2  & 3.62  & 3.78  &  &  & 0.05  & 0.09 &  & \tabularnewline
 & 1000  & 1  & 3.86  & 3.98  & 5.13  & 4.60  & 0.18  & 0.22 & 0.83  & 0.60 \tabularnewline
 &  & 2  & 4.13  & 4.22  &  &  & 0.20  & 0.26 &  & \tabularnewline
 & 2000  & 1  & 4.41  & 4.60  & 5.11  & 4.88  & 0.57  & 0.67 & 0.87  & 0.88 \tabularnewline
 &  & 2  & 4.51  & 4.58  &  &  & 0.51  & 0.58 &  & \tabularnewline
 & 5000  & 1  & 4.71  & 4.75  & 5.03  & 4.95  & 0.77  & 0.80 & 0.97  & 0.95 \tabularnewline
 &  & 2  & 4.83  & 4.84  &  &  & 0.83  & 0.84 &  & \tabularnewline
 & 10000  & 1  & 4.87  & 4.88  & 5.02  & 4.98  & 0.90  & 0.91 & 0.98  & 0.98 \tabularnewline
 &  & 2  & 4.88  & 4.90  &  &  & 0.88  & 0.90 &  & \tabularnewline
4  & 300  & 1  & 2.13  & 2.33  & 5.69  & 2.85  & 0.00  & 0.00 & 0.34  & 0.01 \tabularnewline
 &  & 2  & 2.39  & 2.54  &  &  & 0.00  & 0.00 &  & \tabularnewline
 & 500  & 1  & 2.58  & 2.75  & 5.63  & 3.61  & 0.00  & 0.00 & 0.35  & 0.16 \tabularnewline
 &  & 2  & 3.01  & 3.22  &  &  & 0.00  & 0.00 &  & \tabularnewline
 & 1000  & 1  & 3.44  & 3.70  & 5.61  & 4.75  & 0.03  & 0.09 & 0.61  & 0.76 \tabularnewline
 &  & 2  & 3.82  & 3.94  &  &  & 0.01  & 0.03 &  & \tabularnewline
 & 2000  & 1  & 4.05  & 4.35  & 5.26  & 4.95  & 0.32  & 0.44 & 0.82  & 0.95 \tabularnewline
 &  & 2  & 4.35  & 4.48  &  &  & 0.35  & 0.48 &  & \tabularnewline
 & 5000  & 1  & 4.96  & 4.97  & 5.09  & 5.00  & 0.96  & 0.97 & 0.92  & 1.00 \tabularnewline
 &  & 2  & 4.87  & 4.91  &  &  & 0.87  & 0.91 &  & \tabularnewline
 & 10000  & 1  & 4.98  & 5.00  & 5.03  & 5.00  & 0.99  & 1.00 & 0.97  & 1.00 \tabularnewline
 &  & 2  & 5.00  & 5.00  &  &  & 1.00  & 1.00 &  & \tabularnewline
\hline 
\end{tabular}

\end{table}

\begin{table}[ht]
\caption{{MeanComp and CorrProp results for different values of $\left(d,n_{1},l\right)$,
when $\left(g_{0},\bar{\omega}\right)=\left(10,0.01\right)$.}}

\centering %
\begin{tabular}{|rrr|rrrr|rrrr|}
\hline 
 &  &  &  & MeanComp  &  &  &  & CorrProp &  & \tabularnewline
$d$  & $n_{1}$  & $l$  & Split  & Swapped  & AIC & BIC & Split  & Swapped & AIC & BIC\tabularnewline
\hline 
\hline 
2  & 300  & 1  & 4.80  & 6.12  & 10.13  & 9.12  & 0.01  & 0.01 & 0.45  & 0.37 \tabularnewline
 &  & 2  & 6.71  & 7.48  &  &  & 0.03  & 0.06 &  & \tabularnewline
 & 500  & 1  & 5.90  & 6.81  & 10.14  & 9.33  & 0.01  & 0.03 & 0.44  & 0.43 \tabularnewline
 &  & 2  & 7.98  & 8.33  &  &  & 0.12  & 0.14 &  & \tabularnewline
 & 1000  & 1  & 6.46  & 7.50  & 10.23  & 9.92  & 0.04  & 0.13 & 0.82  & 0.92 \tabularnewline
 &  & 2  & 8.69  & 9.05  &  &  & 0.25  & 0.29 &  & \tabularnewline
 & 2000  & 1  & 7.59  & 8.42  & 10.17  & 9.98  & 0.29  & 0.40 & 0.86  & 0.98 \tabularnewline
 &  & 2  & 9.31  & 9.52  &  &  & 0.47  & 0.58 &  & \tabularnewline
 & 5000  & 1  & 8.76  & 9.14  & 10.07  & 10.00  & 0.69  & 0.78 & 0.94  & 1.00 \tabularnewline
 &  & 2  & 9.50  & 9.79  &  &  & 0.76  & 0.84 &  & \tabularnewline
 & 10000  & 1  & 8.81  & 9.19  & 10.03  & 10.00  & 0.75  & 0.82 & 0.97  & 1.00 \tabularnewline
 &  & 2  & 9.60  & 9.81  &  &  & 0.82  & 0.90 &  & \tabularnewline
4  & 300  & 1  & 3.52  & 3.89  & 10.43  & 8.01  & 0.00  & 0.00 & 0.36  & 0.12 \tabularnewline
 &  & 2  & 4.35  & 4.84  &  &  & 0.00  & 0.00 &  & \tabularnewline
 & 500  & 1  & 4.66  & 5.26  & 10.35  & 8.84  & 0.00  & 0.01 & 0.38  & 0.27 \tabularnewline
 &  & 2  & 5.60  & 6.17  &  &  & 0.00  & 0.01 &  & \tabularnewline
 & 1000  & 1  & 5.83  & 6.50  & 10.47  & 9.85  & 0.00  & 0.00 & 0.68  & 0.85 \tabularnewline
 &  & 2  & 7.17  & 7.77  &  &  & 0.00  & 0.00 &  & \tabularnewline
 & 2000  & 1  & 7.04  & 7.82  & 10.34  & 9.99  & 0.07  & 0.12 & 0.75  & 0.99 \tabularnewline
 &  & 2  & 8.81  & 9.14  &  &  & 0.16  & 0.26 &  & \tabularnewline
 & 5000  & 1  & 8.54  & 9.25  & 10.08  & 9.99  & 0.47  & 0.60 & 0.93  & 0.99 \tabularnewline
 &  & 2  & 9.65  & 9.73  &  &  & 0.73  & 0.76 &  & \tabularnewline
 & 10000  & 1  & 9.37  & 9.68  & 10.06  & 10.00  & 0.84  & 0.91 & 0.95  & 1.00 \tabularnewline
 &  & 2  & 9.87  & 9.93  &  &  & 0.90  & 0.93 &  & \tabularnewline
\hline 
\end{tabular}

\end{table}

\begin{table}[ht]
\caption{{MeanComp and CorrProp results for different values of $\left(d,n_{1},l\right)$,
when $\left(g_{0},\bar{\omega}\right)=\left(10,0.05\right)$.}}

\centering %
\begin{tabular}{|rrr|rrrr|rrrr|}
\hline 
 &  &  &  & MeanComp  &  &  &  & CorrProp &  & \tabularnewline
$d$  & $n_{1}$  & $l$  & Split  & Swapped  & AIC & BIC & Split  & Swapped & AIC & BIC\tabularnewline
\hline 
\hline 
2  & 300  & 1  & 3.46  & 4.18  & 9.72  & 6.46  & 0.00  & 0.00 & 0.24  & 0.01 \tabularnewline
 &  & 2  & 4.07  & 4.60  &  &  & 0.00  & 0.00 &  & \tabularnewline
 & 500  & 1  & 4.10  & 4.91  & 9.88  & 7.15  & 0.00  & 0.00 & 0.35  & 0.03 \tabularnewline
 &  & 2  & 4.95  & 5.46  &  &  & 0.00  & 0.01 &  & \tabularnewline
 & 1000  & 1  & 4.78  & 5.44  & 10.06  & 8.87  & 0.01  & 0.01 & 0.74  & 0.27 \tabularnewline
 &  & 2  & 6.11  & 6.68  &  &  & 0.00  & 0.00 &  & \tabularnewline
 & 2000  & 1  & 5.74  & 6.56  & 9.96  & 9.33  & 0.01  & 0.01 & 0.88  & 0.52 \tabularnewline
 &  & 2  & 6.96  & 7.85  &  &  & 0.02  & 0.11 &  & \tabularnewline
 & 5000  & 1  & 7.21  & 7.82  & 10.02  & 9.74  & 0.06  & 0.11 & 0.98  & 0.78 \tabularnewline
 &  & 2  & 8.59  & 8.91  &  &  & 0.14  & 0.19 &  & \tabularnewline
 & 10000  & 1  & 7.62  & 8.38  & 9.99  & 9.88  & 0.17  & 0.26 & 0.95  & 0.88 \tabularnewline
 &  & 2  & 8.86  & 9.04  &  &  & 0.24  & 0.33 &  & \tabularnewline
4  & 300  & 1  & 2.45  & 2.64  & 10.54  & 4.36  & 0.00  & 0.00 & 0.30  & 0.00 \tabularnewline
 &  & 2  & 2.75  & 2.97  &  &  & 0.00  & 0.00 &  & \tabularnewline
 & 500  & 1  & 2.96  & 3.30  & 10.66  & 5.54  & 0.00  & 0.00 & 0.32  & 0.00 \tabularnewline
 &  & 2  & 3.41  & 3.71  &  &  & 0.00  & 0.00 &  & \tabularnewline
 & 1000  & 1  & 3.89  & 4.32  & 10.45  & 7.89  & 0.00  & 0.00 & 0.68  & 0.05 \tabularnewline
 &  & 2  & 4.98  & 5.30  &  &  & 0.00  & 0.00 &  & \tabularnewline
 & 2000  & 1  & 5.07  & 5.63  & 10.41  & 9.28  & 0.00  & 0.00 & 0.75  & 0.45 \tabularnewline
 &  & 2  & 6.30  & 6.63  &  &  & 0.00  & 0.00 &  & \tabularnewline
 & 5000  & 1  & 6.63  & 7.45  & 10.11  & 9.87  & 0.00  & 0.04 & 0.89  & 0.87 \tabularnewline
 &  & 2  & 8.35  & 8.71  &  &  & 0.01  & 0.05 &  & \tabularnewline
 & 10000  & 1  & 8.25  & 8.57  & 10.04  & 9.97  & 0.13  & 0.18 & 0.96  & 0.97 \tabularnewline
 &  & 2  & 9.23  & 9.36  &  &  & 0.32  & 0.42 &  & \tabularnewline
\hline 
\end{tabular}

\end{table}

\begin{table}[ht]
\caption{\label{tab: 10,0.1}{MeanComp and CorrProp results for different values
of $\left(d,n_{1},l\right)$, when $\left(g_{0},\bar{\omega}\right)=\left(10,0.1\right)$.}}

\centering %
\begin{tabular}{|rrr|rrrr|rrrr|}
\hline 
 &  &  &  & MeanComp  &  &  &  & CorrProp &  & \tabularnewline
$d$  & $n_{1}$  & $l$  & Split  & Swapped  & AIC & BIC & Split  & Swapped & AIC & BIC\tabularnewline
\hline 
\hline 
2  & 300  & 1  & 2.76  & 3.09  & 9.08  & 4.61  & 0.00  & 0.00 & 0.16  & 0.00 \tabularnewline
 &  & 2  & 3.13  & 3.57  &  &  & 0.00  & 0.00 &  & \tabularnewline
 & 500  & 1  & 3.10  & 3.51  & 9.28  & 5.24  & 0.00  & 0.00 & 0.21  & 0.00 \tabularnewline
 &  & 2  & 3.98  & 4.56  &  &  & 0.00  & 0.00 &  & \tabularnewline
 & 1000  & 1  & 4.14  & 4.74  & 9.61  & 6.72  & 0.00  & 0.00 & 0.52  & 0.03 \tabularnewline
 &  & 2  & 5.02  & 5.28  &  &  & 0.00  & 0.00 &  & \tabularnewline
 & 2000  & 1  & 4.99  & 5.55  & 9.79  & 7.93  & 0.00  & 0.00 & 0.77  & 0.08 \tabularnewline
 &  & 2  & 5.87  & 6.30  &  &  & 0.00  & 0.00 &  & \tabularnewline
 & 5000  & 1  & 6.02  & 6.51  & 9.86  & 8.82  & 0.00  & 0.00 & 0.85  & 0.25 \tabularnewline
 &  & 2  & 7.16  & 7.60  &  &  & 0.00  & 0.01 &  & \tabularnewline
 & 10000  & 1  & 6.75  & 7.45  & 9.95  & 9.41  & 0.02  & 0.07 & 0.93  & 0.53 \tabularnewline
 &  & 2  & 7.75  & 8.06  &  &  & 0.04  & 0.06 &  & \tabularnewline
4  & 300  & 1  & 1.89  & 2.01  & 10.44  & 2.77  & 0.00  & 0.00 & 0.26  & 0.00 \tabularnewline
 &  & 2  & 1.92  & 2.17  &  &  & 0.00  & 0.00 &  & \tabularnewline
 & 500  & 1  & 2.27  & 2.53  & 10.53  & 3.66  & 0.00  & 0.00 & 0.28  & 0.00 \tabularnewline
 &  & 2  & 2.57  & 2.73  &  &  & 0.00  & 0.00 &  & \tabularnewline
 & 1000  & 1  & 3.07  & 3.37  & 10.32  & 5.64  & 0.00  & 0.00 & 0.69  & 0.00 \tabularnewline
 &  & 2  & 3.61  & 3.82  &  &  & 0.00  & 0.00 &  & \tabularnewline
 & 2000  & 1  & 4.12  & 4.52  & 10.39  & 7.54  & 0.00  & 0.00 & 0.74  & 0.04 \tabularnewline
 &  & 2  & 4.99  & 5.22  &  &  & 0.00  & 0.00 &   &  \tabularnewline
 & 5000  & 1  & 5.40  & 6.09  & 10.08  & 9.23  & 0.00  & 0.00 & 0.89 & 0.40 \tabularnewline
 &  & 2  & 6.95  & 7.32  &  &  & 0.00  & 0.00 &   &  \tabularnewline
 & 10000  & 1  & 6.90  & 7.60  & 10.06  & 9.77  & 0.01  & 0.03 &  0.97 &  0.79\tabularnewline
 &  & 2  & 8.38  & 8.54  &  &  & 0.02  & 0.02 &  & \tabularnewline
\hline 
\end{tabular}

\end{table}

For the other pairs of $\left(g_{0},\bar{w}\right)$, we observe the
same relationships between the values of $l$ and the Split and Swapped
tests. That is, $l=2$ tends to be more powerful than $l=1$ (except
when $n_{1}$ is relatively small, i.e. $n_{1}\in\left\{{300,500},1000,2000\right\} $),
and the Swapped test tends to be more powerful than the Split test.
In addition, we also observe that the STP becomes more powerful as
$n_{1}$ increases, which supports the conclusions of Theorem \ref{thm: individual test consistency},
which applies to the normal mixture model that is under study. 

For smaller sample sizes, we observe that the STP tended to be more
powerful when $d=2$ in almost all cases, and for larger sample sizes,
the opposite appears to be true. This is likely due to a combination
of the variability of the maximum likelihood estimator and the increase
in separability of higher dimensional spaces. Finally, we notice that
the STP was more powerful when the data were more separable (i.e.,
for smaller values of $\bar{\omega}$). Here, we can see that for $n_{1}=10000$,
the STP can identify the generative value of $g$ in the $g_{0}=5$
scenarios, in a large proportion of cases. However, when $g_{0}=10$,
the STP becomes less powerful. It is particularly remarkable that
even when $n_{1}=10000$, the highest detection proportion was $7\%$
in the $\left(g_{0},\bar{\omega}\right)=\left(10,0.1\right)$ scenarios.
This again implies that the STP lacks power, when applied with the
Split or Swapped tests, especially when component densities of the
generative mixture model are not well separated.

{
Regarding the AIC and BIC point estimators, we observe firstly that across all scenarios, the AIC procedure produces a larger estimate of $g_{0}$ than the BIC procedure, when observing the MeanComp values. We also observe that the AIC estimator is often larger than $g_{0}$, even for larger sample sizes. This observation is in concordance with the theory of \cite{Leroux1992} and \cite{Keribin2000} who show that the AIC procedure does not underestimate $g_{0}$, asymptotically, but is also not consistent. On the other hand, we observe that the BIC procedure tends to underestimate $g_{0}$, for small sample sizes, but becomes more accurate, on average, as $n_{1}$ increases. Again, this is in concordance with the consistency results regarding the BIC estimator of \cite{Keribin2000}.
Regarding the CorrPro values, we observe that in smaller sample sizes ($n_{1}\in\left\{ 300,500\right\}$), the AIC procedure outperforms the BIC procedure in all cases other than those reported in Table \ref{tab: 5,0.01}. This is likely due to the downward bias of the BIC estimates, as observed via the MeanComp values. This downward bias appears to be most apparent in situations where the data are less separable and for larger $g_{0}$, as is evident by the results of Table \ref{tab: 10,0.1}, where the AIC procedure outperforms the BIC procedure with respect to CorrProp, across all sample sizes. In comparison to the STP estimates $\hat{g}_{n}$, as expected, it is notable that the AIC and BIC procedures provide point estimates of $g_{0}$ that are as accurate or more accurate in all simulation scenarios. 
}

Overall, we observe that the conclusions of Theorems \ref{thm: correctness of procedure}--\ref{thm: individual test consistency}
appear to hold over the assessed simulation scenarios. From a practical
perspective we can make the following recommendations. Firstly, the
STP based on the Swapped test is preferred over the Split test. Secondly,
the alternative based on $l=2$ is preferred over $l=1$. Thirdly,
{to obtain intervals of a fixed level of efficiency,} larger sample sizes are necessary when data arise from mixture models
with larger numbers of mixture components and when the mixture components
are not well separated. {Finally, we note that the AIC and BIC procedures both provide accurate point estimation of $g_{0}$ and are both complementary to the interval estimator $\left\{ g_{0}\ge\hat{g}_{n}\right\}$  obtained from the STP.}

\subsection{{Example applications}}

We procedure to demonstrate the utility of the STP via example applications of varying complexity.

\subsubsection{Old Faithful data}

\begin{figure}
\begin{centering}
\includegraphics[width=15cm]{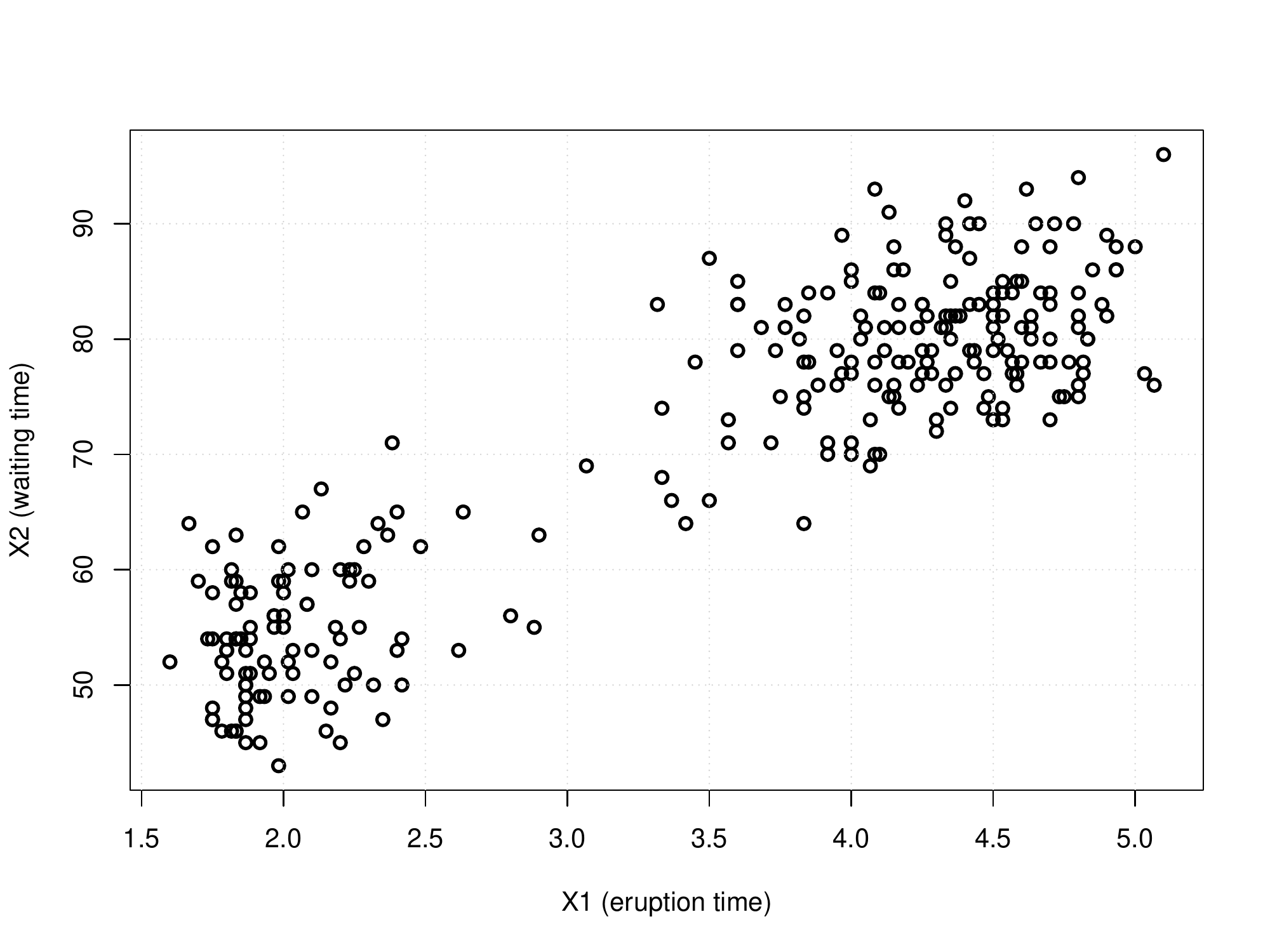}
\par\end{centering}
\caption{\label{fig: faithful}Scatter plot of the $\mathtt{faithful}$ data
set.}
\end{figure}

Our first example is to assess the number of Gaussian mixture components
that are present in the $\mathtt{faithful}$ data set from $\mathsf{R}$,
which was originally studied in \citet{Hardle1991Smoothing-Techn}.
The data set consists of a length $n=272$ realizations of a sequence
$\mathbf{X}_{n}$, where $\bm{X}_{i}\in\mathbb{R}^{2}$ for each $i\in\left[n\right]$.
Here, each observation $\bm{X}_{i}=\left(X_{i1},X_{i2}\right)$ contains
measurements regarding the eruption length of time $X_{i1}$ and the
waiting time until the next eruption $X_{i2}$, in minutes, of eruption
event $i$, for the Old Faithful geyser in Yellowstone National Park,
Wyoming, USA. A visualization of the data appears in Figure \ref{fig: faithful}.

We apply the STP using a $n_{1}=n_{2}=136$ split. The $p$-values
obtained from the Split tests of hypotheses $\text{H}_{g}$ versus
$\bar{\text{H}}_{g}$ with $l_{g}=2$, for $g=1,2$ are $3.40\times10^{-32}$
and $1$. Respectively, the $p$-values for the Swapped test are $6.80\times10^{-32}$
and $1$. Thus, using either the Split or the Swapped test variants
of the STP, for $\alpha>6.80\times10^{-32}$, we can conclude that
the event $\left\{ g_{0}\ge2\right\} $ occurs with a probability
of at least $1-\alpha$. We also obtain the $\text{AIC}_{g}$ values
for $g=1,2,3$: $9.53$, $8.40$, and $8.42$, and the respective
$\text{BIC}_{g}$ values: $9.61$, $8.56$, and $8.66$. Thus, both
procedures estimate the order of the underlying mixture distribution
to be $2$. Although there is no ground truth regarding the $\mathtt{faithful}$
data set, a visual inspection of Figure \ref{fig: faithful} suggests
that both the STP interval estimator and the point estimation provided
by the AIC and BIC procedures are reasonable.

\subsubsection{Palmer penguins data}

\begin{figure}
\begin{centering}
\includegraphics[width=15cm]{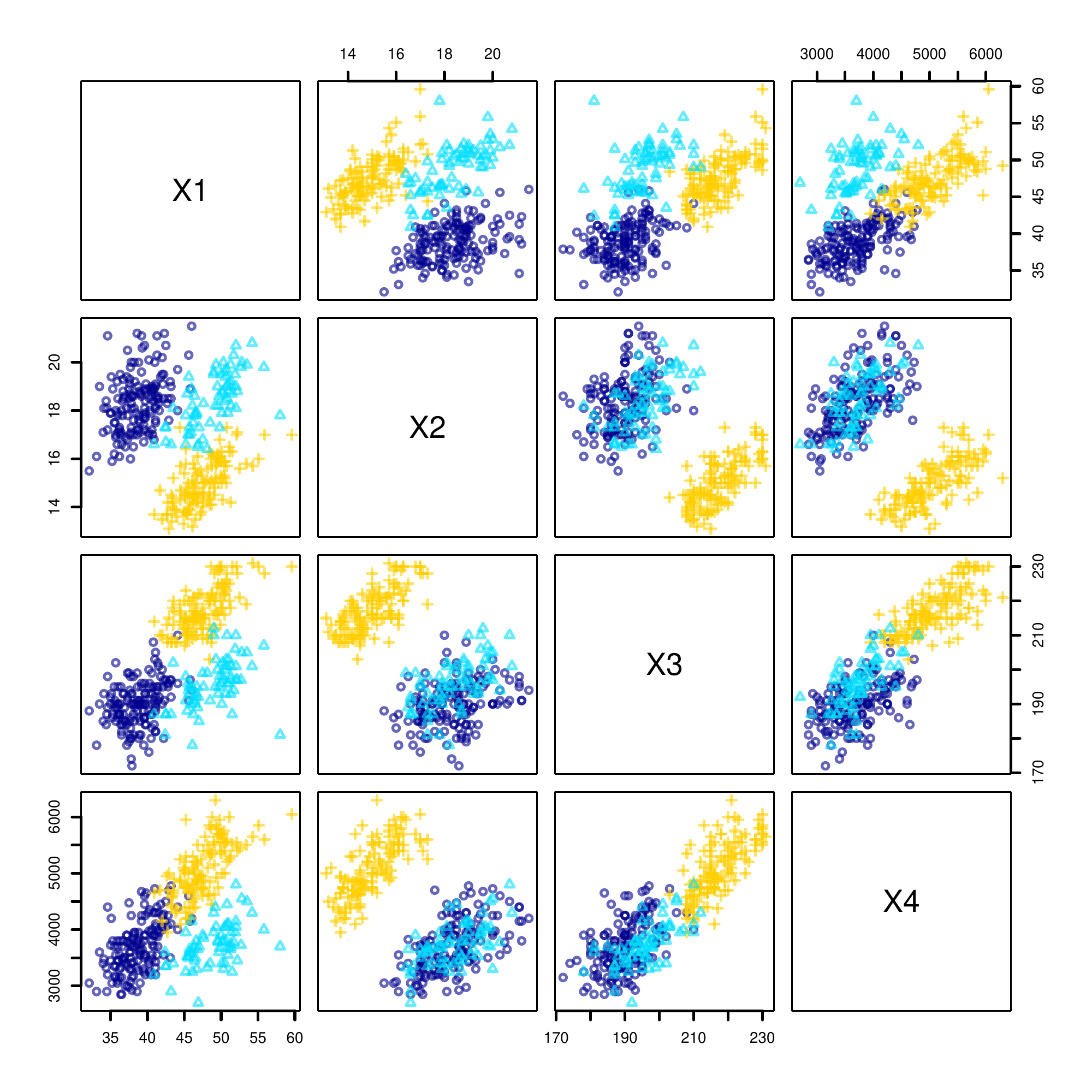}
\par\end{centering}
\caption{\label{fig: penguins}Scatter plot of the $\mathtt{penguins}$ data
set. Adelie, Chinstrap, and Gentoo data points are plotted as circles,
triangles, and plus signs, respectively.}
\end{figure}

Our second example is to estimate the Gaussian mixture order of the
$\mathtt{penguins}$ data set from the $\mathsf{R}$ package $\mathsf{palmerpenguins}$,
originally considered by \citet{Gorman2014Ecological-sexu}. After
removing rows with missing data, the data set contains a length $n=342$
realization of a sequence $\mathbf{X}_{n}$, where $\bm{X}_{i}\in\mathbb{R}^{4}$
for each $i\in\left[n\right]$. Here, each observation $\bm{X}_{i}=\left(X_{i1},X_{i2},X_{i3},X_{i4}\right)$
contains measures regarding penguins of the Adelie, Gentoo, and Chinstrap
species. Specifically, for each $i$, the measurements are the bill
length $X_{i1}$, bill depth $X_{i2}$, and flipper length $X_{i3}$,
all in millimeters, along with the body mass $X_{i4}$, in grams.
A visualization of the data, with separate symbols for the different
penguin species, is provided in Figure \ref{fig: penguins}.

We apply the STP using a $n_{1}=n_{2}=171$ split. The $p$-values
obtained from the Split tests of hypotheses $\text{H}_{g}$ versus
$\bar{\text{H}}_{g}$ with $l_{g}=2$, for $g=1,2$ are $3.78\times10^{-61}$
and $1$. Respectively, the $p$-vales for the Swapped test are $9.25\times10^{-66}$
and $1$. Thus, using either version of the STP, for $\alpha>3.78\times10^{-61}$,
we conclude that the event $\left\{ g_{0}\ge2\right\} $ occurs with
a probability of at least $1-\alpha$. For these data, the $\text{AIC}_{g}$
values for $g=1,2,3,4,5$ are $32.37$, $30.65$, $30.38$, $30.35$,
and $30.36$., and the respective $\text{BIC}_{g}$ values are $32.54$,
$30.99$, $30.89$, $31.03$, and $31.20$. Thus, the AIC and BIC
procedures estimate the true order $g_{0}$ to be $4$ and $3$, respectively.
Compared to the ground truth of three penguin species, we observe
that the AIC procedure is an over estimate, whereas the BIC is accurate.
The inference obtained from the STP is also correct, with the assessment
that there are at least $2$ mixture components, with high probability.

\subsubsection{Cell lines data set}

\begin{figure}
\begin{centering}
\includegraphics[width=15cm]{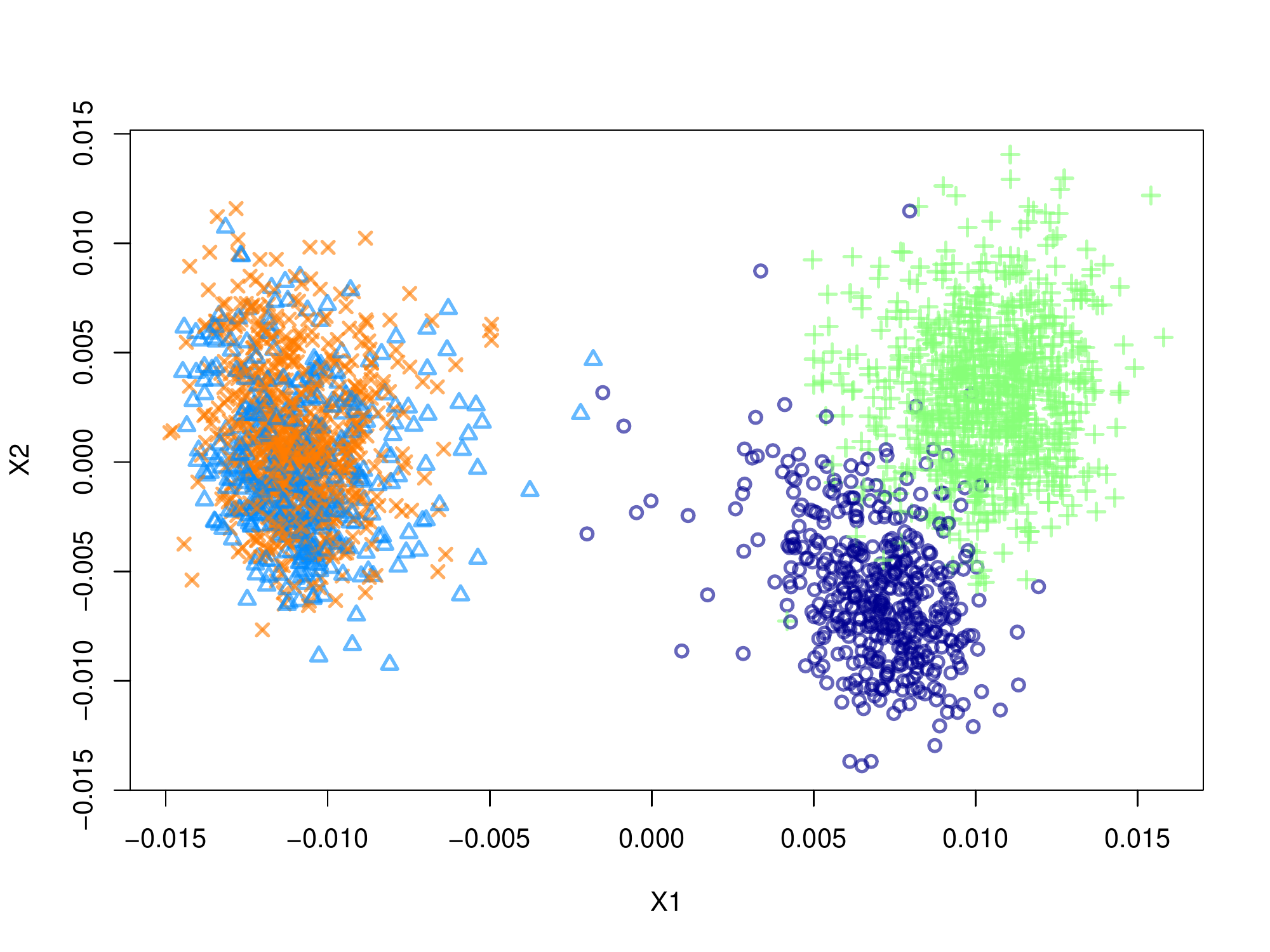}
\par\end{centering}
\caption{\label{fig: cell_lines}Scatter plot of the $\texttt{cell\_lines}$
data set. The pure and mixed Jurkat cells data are plotted as plus
signs and circles, respectively, and the pure and mixed HEK293T data
are plotted as crosses and triangles, respectively.}
\end{figure}

Our final example is to identify the number of mixture components
in the $\texttt{cell\_lines}$ data set from the $\mathsf{harmony}$
package of \citet{Korsunsky2019Fast-sensitive-}. As presented in
\url{https://portals.broadinstitute.org/harmony/articles/quickstart.html},
the data set consists of $n=2370$ rows consisting of a realization
of the sequence $\mathbf{X}_{n}$ of random variable $\bm{X}_{i}\in\mathbb{R}^{2}$,
for each $i\in\left[n\right]$. Each observation $\bm{X}_{i}=\left(X_{i1},X_{i2}\right)$
contains measurements of the first and second scaled principal components
of single cell gene expression data. The data come from three sources,
where the first source comes form a pure Jurkat cell lines, the second
comes from a pure HEK293T cell lines, and the third source consists
of a half-and-half mix of Jurkat cells and HEK293T cells. Since the
data from the mixed sources are not registered to the pure sources
data, there are in effect four separate subpopulations of observations.
We plot the $\texttt{cell\_lines}$ data in Figure \ref{fig: cell_lines}.

We apply the STP using a $n_{1}=n_{2}=1185$ split. The $p$-values
obtained from the Split tests of hypotheses $\text{H}_{g}$ versus
$\bar{\text{H}}_{g}$ with $l_{g}=2$, for $g=1,2,3,4$, are $0$
(in double precision zero), $5.22\times10^{-49}$, $5.70\times10^{-13}$,
and $0.21$. Respectively, the Swapped tests yield $p$-values $0$,
$1.04\times10^{-48}$, $2.39\times10^{-17}$, and $0.42$. Thus, for
any $\alpha>5.70\times10^{-13}$, the STP concludes that the event
$\left\{ g_{0}\ge4\right\} $ occurs with a probability of at least
$1-\alpha$. Again, we compute the $\text{AIC}_{g}$ and $\text{BIC}_{g}$
values. For each $g=1,2,3,4,5,6,7,8$, the $\text{AIC}_{g}$ values
are $-14.23$, $-16.32$, $-16.45$, $-16.52$, $-16.54$, $-16.55$,
$-16.56$, and $-16.55$, respectively. Thus, The AIC procedure estimates
$g_{0}$ as 7. For each $g=1,2,3,4,5,6$, the $\text{BIC}_{g}$ values
are $-14.22$, $-16.29$, $-16.41$, $-16.46$, $-16.47$, and $-16.46$.
Thus, the BIC procedure estimates the mixture order to be $5$. Compared
to the ground truth, it appears that the confidence set $\left\{ g_{0}\ge4\right\} $
provides sensible inference regarding the underlying number of Gaussian
mixture components. It would appear that the AIC and BIC procedures
both overestimate the underling mixture order. However, it could
also be true that the subpopulations corresponding to each of the
cell lines cannot be adequately modeled via Gaussian mixture components.

\section{Extensions}

\subsection{A consistent sequential testing procedure}

Important criteria regarding the validity of an order selection
method are the large sample properties of conservativeness and consistency.
These properties are defined by \citet{Leeb:2009aa}, in the context
of this work, as
\[
\lim_{n_{1},n_{2}\rightarrow\infty}\mathrm{Pr}_{f_{0}}\left(g_{0}\ge\hat{g}_{n}\right)=1
\]
and
\[
\lim_{n_{1},n_{2}\rightarrow\infty}\mathrm{Pr}_{f_{0}}\left(g_{0}=\hat{g}_{n}\right)=1\text{,}
\]
for all $f_{0}\in\mathcal{M}$, respectively (see also \citealp[Sec. 7.1]{Dickhaus2014}).

By Theorem \ref{thm: correctness of procedure}, we have the fact
that (\ref{eq: confidence statement}) holds for all $n$, and thus
the STP, as stated in Section \ref{sec:Introduction},
cannot be conservative, nor consistent. However, if we replace $\alpha$
by a sequence $\left(\alpha_{n}\right)_{n=1}^{\infty}$, where $\alpha_{n}\rightarrow0$
as $n_{1},n_{2}\rightarrow\infty$, then we can conclude that the
modified procedure is conservative by taking the limits on both sides
of inequality (\ref{eq: confidence statement}).

We now specialize our focus, again, to the parametric setting. To construct a procedure that is consistent requires further modification
to the STP. Namely, we require additionally
that the individual tests of $\text{H}_{g}$ are consistent (i.e.,
that Theorem \ref{thm: individual test consistency} holds for the
sequence $\left(\alpha_{n}\right)_{n=1}^{\infty}$, replacing $\alpha$
in each test). Thus, to make (\ref{eq: criterion}) hold with probability
approaching one, we require that the third term on the left-hand side
converges to zero. We observe that the sequence $\left(\alpha_{n}\right)_{n=1}^{\infty}$
must simultaneously satisfy the conditions that $\alpha_{n}\rightarrow0$
and $n_{1}^{-1}\log\alpha_{n}\rightarrow0$, as $n_{1},n_{2}\rightarrow\infty$.
For instance, we may choose to set $\alpha_{n}=n_{1}^{-\kappa}$,
with $\kappa>0$. We thus have the following result regarding the
STP when applied using the sequence of $p\text{-values}$
$\left(P_{g}^{1}\left(\mathbf{X}_{n}\right)\right)_{g=1}^{\infty}$.
\begin{cor} \label{cor: Consistent model selection}
Assume (A1)--(A4) from Theorem \ref{thm: individual test consistency},
and that $g_{0}<\infty$. If $\alpha_{n}\rightarrow0$ and $n_{1}^{-1}\log\alpha_{n}\rightarrow0$,
as $n_{1},n_{2}\rightarrow\infty$, then the STP
for testing the sequence $\left(\text{H}_{g}\right)_{g=1}^{\infty}$
is consistent, when applied using the rules $\left(R_{g}\left(\mathbf{X}_{n}\right)\right)_{g=1}^{\infty}$,
where $R_{g}\left(\mathbf{X}_{n}\right)=\mathbf{1}\left(P_{g}^{1}\left(\mathbf{X}_{n}\right)<\alpha_{n}\right)$.
\end{cor}
\begin{proof}
The proof of this result appears in the Appendix.
\end{proof}
We note that the modified STP resembles the
time series order selection procedure of \citet{Potscher:1983aa}.
In fact, the conditions placed on the sequence $\left(\alpha_{n}\right)_{n=1}^{\infty}$
are the same as those imposed in \citet[Thm. 5.7]{Potscher:1983aa}.
Furthermore, we note that the conditions placed on $\left(\alpha_{n}\right)_{n=1}^{\infty}$
closely resemble the conditions that are required for the consistent
application of information criteria methods; see
\citet{Keribin2000} and \citet{Baudry:2015aa}. {
We can observe this resemblance by considering expression (\ref{eq: BIC def}) and
taking $\text{BIC}_{g+l}-\text{BIC}_{g}$, for any $g,l\in\mathbb{N}$.
For the BIC procedure to be consistent, this expression must be negative,
for large $n$, which requires that the difference in penalty $n^{-1}\log n\left(\text{dim}_{g+1}-\text{dim}_{g}\right)$
goes to zero. In the STP, if we set $n_{1}=n/2$ (or as any fraction of $n$) and $\alpha_{n}=1/n$, then we have the similar
requirement (of the same rate in $n$) that $\left(2n\right)^{-1}\log n$
must go to zero.
}

\subsection{Asymptotic tests}

Throughout the manuscript, we have assumed that the $p\text{-values}$
from which tests are constructed satisfy (\ref{eq: P-value def})
for all $n$. This assumption is compatible with our application of
the STP using the local tests proposed in
Section \ref{sec:Local-order-testing}. We note that the STP still provides guarantees
for $p\text{-values}$ that only satisfy (\ref{eq: P-value def})
asymptotically, in the sense that 
\begin{equation}
\limsup_{n\rightarrow\infty}\text{Pr}_{f}\left(P_{g}\left(\mathbf{X}_{n}\right)\le\alpha\right)\le\alpha\label{eq: P-value asy}
\end{equation}
for all $f\in\mathcal{M}_{g}$. In such a case, we have the limiting
version of the confidence statement (\ref{eq: confidence statement}):

\begin{equation}
\liminf_{n\rightarrow\infty}\text{Pr}_{f_{0}}\left(g_{0}\ge\hat{g}_{n}\right)\ge1-\alpha\text{.}\label{eq: liminf version main}
\end{equation}

To obtain (\ref{eq: liminf version main}), suppose that $f_{0}\in\mathcal{M}_{g_{0}}$,
for some finite $g_{0}\in\mathbb{N}$. In the notation of Section
\ref{sec:Confidence-via-the}, we can write $\mathbb{G}_{0}\left(f_{0}\right)=\mathbb{N}\backslash\left[g_{0}-1\right]$,
and hence
\begin{align}
\text{Pr}_{f_{0}}\left(f_{0}\in\mathcal{M}_{\hat{g}_{n}-1}\right) & =\mathrm{Pr}_{f_{0}}\left(\bigcup_{g\in\mathbb{N}\backslash\left[g_{0}-1\right]}\left\{ \bar{R}_{g}\left(\mathbf{X}_{n}\right)=1\right\} \right)=\mathrm{Pr}_{f_{0}}\left(\bar{R}_{g_{0}}\left(\mathbf{X}_{n}\right)=1\right)\nonumber \\
 & =\mathrm{Pr}_{f_{0}}\left(\bigcap_{g\in\left[g_{0}\right]}\left\{ P_{g}\left(\mathbf{X}_{n}\right)\le\alpha\right\} \right)\le\mathrm{Pr}_{f_{0}}\left(P_{g_{0}}\left(\mathbf{X}_{n}\right)\le\alpha\right)\text{,}\label{eq: fwer expanded}
\end{align}
Then, since (\ref{eq: fwer expanded}) holds for all $n$, we can
apply \citet[Thm. 3.19]{Rudin:1976aa} to obtain
\[
\limsup_{n\rightarrow\infty}\text{Pr}_{f_{0}}\left(f_{0}\in\mathcal{M}_{\hat{g}_{n}-1}\right)\le\limsup_{n\rightarrow\infty}\mathrm{Pr}_{f_{0}}\left(P_{g_{0}}\left(\mathbf{X}_{n}\right)\le\alpha\right)\le\alpha\text{,}
\]
as required. Using (\ref{eq: liminf version main}), we can justify
the use of the STP with asymptotically valid
tests, such as the procedure of \citet{Li:2010aa}.

\subsection{Aggregated tests}

Under the null hypothesis that $f_{0}\in\mathcal{M}_{g}$, both the
Split and Swapped statistics, $V_{g}^{k}\left(\mathbf{X}_{n}\right)$
and $\bar{V}_{g}\left(\mathbf{X}_{n}\right)$, are examples of $e$-values
(which we shall write generically as $E_{g}$), as defined in \citet{Vovk:2021wm}
(note that these values also appear as $s$-values in \citealp{Grunwald:2020vf},
and as betting scores in \citealp{Shafer:2021vh}), based on the defining
feature that
\begin{equation}
\sup_{f\in\mathcal{M}_{g}}\text{E}_{f}\left(E_{g}\right)\le1\text{.}\label{eq: e-value}
\end{equation}
By Markov's inequality, (\ref{eq: e-value}) implies
\[
\sup_{f\in\mathcal{M}_{g}}\text{Pr}_{f}\left(E_{g}\ge1/\alpha\right)\le\alpha\text{,}
\]
for any $\alpha\in\left(0,1\right)$, from which we can derive the
$p$-value $\max\{P_{g}=1/E_{g},1\}$, which satisfies (\ref{eq: P-value def}).

As discussed in \citet{Wasserman:2020aa},
any set of possibly dependent $e$-values $E_{g}^{1},\dots,E_{g}^{m}$
($m\in\mathbb{N}$) can be combined by simple averaging to generate
a new $e$-value $\bar{E}_{g}=m^{-1}\sum_{j=1}^{m}E_{g}^{j}$, which
we shall call the aggregated $e$-value. As such, one may consider
generating $m$ different $e$-values based on either the Split or
Swapped statistics, using different partitions of the data into subsequences
$\mathbf{X}_{n}^{1}$ and $\mathbf{X}_{n}^{2}$. For any fixed $n_{1}$
and $n_{2}$, there are only a finite number of such partitions and
thus one may imagine an aggregated $e$-value that averages over all
such partitions. This hypothetical process was referred to as derandomization
in \citet{Wasserman:2020aa}, since the resulting $p$-value is no
longer dependent on any particular random partitioning of $\mathbf{X}_{n}$.

We further note that one can also aggregate the results from multiple
instances of the Split and Swapped statistics via methods for aggregating
over $p$-values. These methods are discussed at length in the works
of \citet{Vovk:2020wp} {who provide a detailed assessment of methods for combining arbitrarily dependent $p$-values, via generalized averaging operations}.

\section{Conclusions}

In this work, we proved that the closed testing principle could be
used to construct a sequence of null hypothesis tests that generates
a confidence statement regarding the true number of mixture components
of a finite mixture model. Further, we derive tests for each of the
null hypotheses in the STP, using the universal inference framework
of \citet{Wasserman:2020aa}, and proved that in the parametric case,
under regularity conditions, such tests are consistent against fixed
alternative hypotheses.

The performance of the STP for order selection of normal mixture models
was considered via a comprehensive simulation study. We observe from
the study that the constructed confidence statements were conservative,
as predicted by the theory, and we were also able to make recommendations
regarding the different variants of the tests, for practical application. {We also determined that the AIC and BIC point estimators provide accurate complements to the intervals provided by the STP. Example applications of the STP are further described to demonstrate the utility of our methods in practice.} {We recommend that our STP interval estimators be reported alongside an AIC or BIC point estimator to provide both an accurate and precise inference regarding the true order.}

Extensions of the STP were also discussed, including the possibility
of aggregating over multiple tests, and performing the STP with asymptotic
tests. Of particular interest is a proof that the testing procedure
could be modified to generate an order selection procedure that
consistently determines the true number of mixture components,
in the asymptotic sense. Our proof shows that such a procedure was
essentially equivalent to other asymptotic model selection methods
such as the Bayesian information criterion and its variants.

We note that our general order selection confidence result of Theorem
\ref{thm: correctness of procedure} applies not only to finite mixture
models, but also to any nested sequences of models. For example, we
may consider the same STP to generate confidence statements regarding
the number of factors in a factor analysis model or the degree of
a polynomial fit. We leave the application of the STP to such problems
for future work, along with the applications of our discussed variants
on the testing procedures.

\section*{{Appendix}}

\subsection*{Proof of Theorem \ref{thm: correctness of procedure}}

Firstly, since $\mathcal{M}_{g}\subset\mathcal{M}_{g+1}$, we have
the fact that for any $g\in\mathbb{I}\subset\mathbb{N}$, $\bigcap_{g\in\mathbb{I}}\mathcal{M}_{g}=\mathcal{M}_{\min_{g\in\mathbb{I}}g}$
and thus the sequence $\left(\text{H}_{g}\right)_{g=1}^{\infty}$
is $\cap\text{-closed}$. Next, the sequential procedure rejects $\text{H}_{g}$
if and only if $R_{j}\left(\mathbf{X}_{n}\right)=1$ for each $j\in\left[g\right]$, or more
compactly, $\text{H}_{g}$ is rejected if and only if $\bar{R}_{g}\left(\mathbf{X}_{n}\right)=\min_{j\in\left[g\right]}R_{j}\left(\mathbf{X}_{n}\right)=1$.
Because $\mathcal{M}_{g}\subset\mathcal{M}_{g+1}$, we also have the
fact that $\left\{ j:\mathcal{M}_{j}\subseteq\mathcal{M}_{g}\right\} =\left[g\right]$,
and thus $\left(\bar{R}_{g}\left(\mathbf{X}_{n}\right)\right)_{g=1}^{\infty}$ is exactly the
sequence of closed tests for $\left(\text{H}_{g}\right)_{g=1}^{\infty}$,
of form (\ref{eq: closed tests}).

By Theorem \ref{thm: closed tests}, for each $f\in\mathcal{M}$,
we have the inequality
\begin{equation}
\mathrm{Pr}_{f}\left(\bigcup_{g\in\mathbb{G}_{0}\left(f\right)}\left\{ \bar{R}_{g}\left(\mathbf{X}_{n}\right)=1\right\} \right)\le\alpha\text{,}\label{eq: generic fwer}
\end{equation}
where the event $\left\{ \bigcup_{g\in\mathbb{G}_{0}\left(f\right)}\left\{ \bar{R}_{g}\left(\mathbf{X}_{n}\right)=1\right\} \right\} $
can be written as $\left\{ f\in\bigcup_{j\in\left[\hat{g}_{n}-1\right]}\mathcal{M}_{j}\right\} $,
since the sequential procedure first fails to reject hypothesis $\text{H}_{\hat{g}_{n}}$.
Again, since $\mathcal{M}_{g}\subset\mathcal{M}_{g+1}$, $\bigcup_{j\in\left[\hat{g}_{n}-1\right]}\mathcal{M}_{j}=\mathcal{M}_{\hat{g}_{n}-1}$
and thus (\ref{eq: generic fwer}) can be written in form (\ref{eq: wasserman statement}).
This completes the proof.

\subsection*{Proof of Theorem \ref{thm: individual test consistency}}

Write the event $\left\{ P_{g}^{1}\left(\mathbf{X}_{n}\right)<\alpha\right\} $
as
\[
\frac{\prod_{i=1}^{n_{1}}f\left(\bm{X}_{i}^{1};\tilde{\bm{\vartheta}}_{n}^{\left(g\right)}\right)}{\prod_{i=1}^{n_{1}}f\left(\bm{X}_{i}^{1};\hat{\bm{\vartheta}}_{n}^{\left(g+l_{g}\right)}\right)}<\alpha\text{,}
\]
or equivalently
\begin{equation}
\frac{1}{n_{1}}\sum_{i=1}^{n_{1}}\log f\left(\bm{X}_{i}^{1};\tilde{\bm{\vartheta}}_{n}^{\left(g\right)}\right)-\frac{1}{n_{1}}\sum_{i=1}^{n_{1}}\log f\left(\bm{X}_{i}^{1};\hat{\bm{\vartheta}}_{n}^{\left(g+l_{g}\right)}\right)-\frac{\log\alpha}{n_{1}}<0\text{.}\label{eq: criterion}
\end{equation}
Thus, it suffices to show that the left-hand side converges in probability
to a constant that is bounded above by zero. 

By (A2) and (A3), we have the facts that (i): $\tilde{\bm{\vartheta}}_{n}^{\left(g\right)}\rightarrow\bm{\vartheta}_{0}^{\left(g\right)}$,
in probability as $n_{1}\rightarrow\infty$, where

\[
\bm{\vartheta}_{0}^{\left(g+l_{g}\right)}\in\left\{ \tilde{\bm{\vartheta}}^{\left(g+l_{g}\right)}\in\mathbb{T}_{g+l_{g}}:\mathrm{E}_{f_{0}}\log f\left(\bm{X};\tilde{\bm{\vartheta}}^{\left(g+l_{g}\right)}\right)=\max_{\bm{\vartheta}^{\left(g+l_{g}\right)}\in\mathbb{T}_{g+l_{g}}}\mathrm{E}_{f_{0}}\log f\left(\bm{X};\bm{\vartheta}^{\left(g+l_{g}\right)}\right)\right\} \text{,}
\]
 and (ii):
\[
\frac{1}{n_{1}}\sum_{i=1}^{n_{1}}\log f\left(\bm{X}_{i}^{1};\tilde{\bm{\vartheta}}_{n}^{\left(g\right)}\right)\rightarrow\text{E}_{f_{0}}\log f\left(\bm{X};\bm{\vartheta}_{0}^{\left(g\right)}\right)\text{,}
\]
in probability, as $n_{1}\rightarrow\infty$, by application of \citet[Lem. 1]{Atienza2007},
which states that
\begin{equation}
\left|\log f\left(\bm{X};\bm{\vartheta}^{\left(g\right)}\right)\right|\le\sum_{z=1}^{g}\left|\log f\left(\bm{X};\bm{\theta}_{z}\right)\right|\text{,}\label{eq: bounding likelihood}
\end{equation}
and using the classic uniform weak law of large numbers of \citet[Thm. 2]{Jennrich1969}.
That is, (A2) permits the use of \citet[Lem. 4.2]{Potscher:1997aa}
to prove result (i), by verifying the conditions for the uniform law,
which can be done via the bound (\ref{eq: bounding likelihood}) and
the existence of moments from (A3). Next, using (i), we show (ii)
by considering the decomposition:
\begin{align*}
\left|\frac{1}{n_{1}}\sum_{i=1}^{n_{1}}\log f\left(\bm{X}_{i}^{1};\tilde{\bm{\vartheta}}_{n}^{\left(g\right)}\right)-\text{E}_{f_{0}}\log f\left(\bm{X};\bm{\vartheta}_{0}^{\left(g\right)}\right)\right| & \le\left|\frac{1}{n_{1}}\sum_{i=1}^{n_{1}}\log f\left(\bm{X}_{i}^{1};\tilde{\bm{\vartheta}}_{n}^{\left(g\right)}\right)-\text{E}_{f_{0}}\log f\left(\bm{X};\tilde{\bm{\vartheta}}_{n}^{\left(g\right)}\right)\right|\\
 & +\left|\text{E}_{f_{0}}\log f\left(\bm{X};\tilde{\bm{\vartheta}}_{n}^{\left(g\right)}\right)-\text{E}_{f_{0}}\log f\left(\bm{X};\bm{\vartheta}_{0}^{\left(g\right)}\right)\right|\text{,}
\end{align*}
where the first term on the right-hand side converges to zero in probability,
by the uniform law, and using (A3), the second term is bounded from
above by 
\begin{equation}
\text{E}_{f_{0}}\left|\log f\left(\bm{X};\tilde{\bm{\vartheta}}_{n}^{\left(g\right)}\right)-\log f\left(\bm{X};\bm{\vartheta}_{0}^{\left(g\right)}\right)\right|\le2g\text{E}_{f_{0}}M_{1}\left(\bm{X}\right)<\infty\text{.}\label{eq: bounding log like diff}
\end{equation}
The continuity from (A2) and bound (\ref{eq: bounding log like diff})
then implies that the second term is continuous with respect to the
argument $\tilde{\bm{\vartheta}}_{n}^{\left(g\right)}$ (cf. \citealp[Thm. 7.1.3]{MakarovPodkorytov2013}).
The continuous mapping theorem then implies that the second term converges
in probability to zero, as $n_{1}\rightarrow\infty$.

Next, we write 
\begin{align*}
&\left|\frac{1}{n_{1}}\sum_{i=1}^{n_{1}}\log f\left(\bm{X}_{i}^{1};\hat{\bm{\vartheta}}_{n}^{\left(g+l_{g}\right)}\right)-\text{E}_{f_{0}}\log f\left(\bm{X};\bm{\vartheta}_{0}^{\left(g+l_{g}\right)}\right)\right|\\ & \le\left|\frac{1}{n_{1}}\sum_{i=1}^{n_{1}}\log f\left(\bm{X}_{i}^{1};\hat{\bm{\vartheta}}_{n}^{\left(g+l_{g}\right)}\right)-\frac{1}{n_{1}}\sum_{i=1}^{n_{1}}\log f\left(\bm{X}_{i}^{1};\bm{\vartheta}_{0}^{\left(g+l_{g}\right)}\right)\right|\\
 & +\left|\frac{1}{n_{1}}\sum_{i=1}^{n_{1}}\log f\left(\bm{X}_{i}^{1};\bm{\vartheta}_{0}^{\left(g+l_{g}\right)}\right)-\text{E}_{f_{0}}\log f\left(\bm{X};\bm{\vartheta}_{0}^{\left(g+l_{g}\right)}\right)\right|\text{.}
\end{align*}
Using (A3), the first term on the right-hand side can be bounded from
above by
\[
\frac{1}{n_{1}}\sum_{i=1}^{n_{1}}\left|\log f\left(\bm{X}_{i}^{1};\hat{\bm{\vartheta}}_{n}^{\left(g+l_{g}\right)}\right)-\log f\left(\bm{X}_{i}^{1};\bm{\vartheta}_{0}^{\left(g+l_{g}\right)}\right)\right|\le\frac{1}{n_{1}}\sum_{i=1}^{n_{1}}M_{2}\left(\bm{X}_{i}^{1}\right)\left\Vert \hat{\bm{\vartheta}}_{n}^{\left(g+l_{g}\right)}-\bm{\vartheta}_{0}^{\left(g+l_{g}\right)}\right\Vert \text{.}
\]
Thus, the first term converges to zero in probability, as $n_{1}\rightarrow\infty$,
by the law of large numbers (since $\text{E}_{f_{0}}M_{2}\left(\bm{X}\right)<\infty$),
and since $\hat{\bm{\vartheta}}_{n}^{\left(g+l_{g}\right)}\rightarrow\bm{\vartheta}_{0}^{\left(g+l_{g}\right)}$,
in probability, as $n_{2}\rightarrow\infty$. The second term converges
to zero, in probability, as $n_{1}\rightarrow\infty$, by the law
of large numbers, since 
\[
\text{E}_{f_{0}}\left|\log f\left(\bm{X};\bm{\vartheta}_{0}^{\left(g+l_{g}\right)}\right)\right|\le2\left(g+l_{g}\right)\text{E}M_{1}\left(\bm{X}\right)<\infty\text{,}
\]
by application of bound (\ref{eq: bounding likelihood}). 

We have thus established that the left-hand side of (\ref{eq: criterion})
converges in probability to 
\begin{equation}
\text{E}_{f_{0}}\log f\left(\bm{X};\bm{\vartheta}_{0}^{\left(g\right)}\right)-\text{E}_{f_{0}}\log f\left(\bm{X};\bm{\vartheta}_{0}^{\left(g+l_{g}\right)}\right)=\text{D}\left(f_{0},f\left(\cdot;\bm{\vartheta}_{0}^{\left(g+l_{g}\right)}\right)\right)-\text{D}\left(f_{0},f\left(\cdot;\bm{\vartheta}_{0}^{\left(g\right)}\right)\right)\text{,}\label{eq: criterion expected}
\end{equation}
as $n_{1},n_{2}\rightarrow\infty$. Suppose, for contradiction, that
(\ref{eq: criterion expected}) is equal to zero. Then, for all $f\left(\bm{x};\bm{\vartheta}^{\left(g+l_{g}\right)}\right)\in\mathcal{M}_{g+l_{g}}$,
\[
\text{D}\left(f_{0},f\left(\cdot;\bm{\vartheta}^{\left(g+l_{g}\right)}\right)\right)-\text{D}\left(f_{0},f\left(\cdot;\bm{\vartheta}_{0}^{\left(g\right)}\right)\right)\ge0\text{.}
\]
 In particular, for some $\bm{\theta}\in\mathbb{T}$ and $\varpi\in\left(0,1\right)$,
we have

\[
\int_{\mathbb{X}}f_{0}\left(\bm{x}\right)\log\left\{ \frac{\left(1-\varpi\right)f\left(\bm{x};\bm{\vartheta}_{0}^{\left(g\right)}\right)+\varpi f\left(\bm{x};\bm{\theta}\right)}{f\left(\bm{x};\bm{\vartheta}_{0}^{\left(g\right)}\right)}\right\} \text{d}\bm{x}\le0\text{.}
\]
By Fatou's Lemma, 
\begin{align*}
0 & \ge\int_{\mathbb{X}}f_{0}\left(\bm{x}\right)\underset{\varpi\rightarrow0}{\lim\inf}\frac{1}{\varpi}\log\left\{ \frac{\left(1-\varpi\right)f\left(\bm{x};\bm{\vartheta}_{0}^{\left(g\right)}\right)+\varpi f\left(\bm{x};\bm{\theta}\right)}{f\left(\bm{x};\bm{\vartheta}_{0}^{\left(g\right)}\right)}\right\} \text{d}\bm{x}\\
 & =\int_{\mathbb{X}}f_{0}\left(\bm{x}\right)\left\{ \frac{f\left(\bm{x};\bm{\theta}\right)}{f\left(\bm{x};\bm{\vartheta}_{0}^{\left(g\right)}\right)}-1\right\} \text{d}\bm{x}\text{,}
\end{align*}
which implies that 
\begin{equation}
\int_{\mathbb{X}}f_{0}\left(\bm{x}\right)\frac{f\left(\bm{x};\bm{\theta}\right)}{f\left(\bm{x};\bm{\vartheta}_{0}^{\left(g\right)}\right)}\text{d}\bm{x}\le1\text{.}\label{eq: ratio}
\end{equation}

Since $f_{0}\in\mathcal{M}\backslash\mathcal{M}_{g}$, we have $f_{0}=f\left(\cdot;\bm{\vartheta}_{0}^{\left(g_{0}\right)}\right)\in\mathcal{M}_{g_{0}}$,
where $g_{0}>g$ and $\bm{\vartheta}^{g_{0}}$ contains the pairs
$\left(\pi_{0,z},\bm{\theta}_{0,z}\right)_{z=1}^{g_{0}}$. By taking
the expectation of both sides of (\ref{eq: ratio}) with respect to
the probability measure on $\bm{\theta}$, defined by
\[
\text{Pr}\left(\bm{\theta}=\bm{\theta}^{\prime}\right)=\sum_{z=1}^{g_{0}}\pi_{0,z}\mathbf{1}\left(\bm{\theta}^{\prime}=\bm{\theta}_{0,z}\right)\text{,}
\]
we have
\[
\sum_{z=1}^{g_{0}}\int_{\mathbb{X}}f_{0}\left(\bm{x}\right)\frac{\pi_{0,z}f\left(\bm{x};\bm{\theta}_{0,z}\right)}{f\left(\bm{x};\bm{\vartheta}_{0}^{\left(g\right)}\right)}\text{d}\bm{x}=\int_{\mathbb{X}}\frac{f_{0}^{2}\left(\bm{x}\right)}{f\left(\bm{x};\bm{\vartheta}_{0}^{\left(g\right)}\right)}\text{d}\bm{x}\le1\text{.}
\]
Finally, by the fact that $\log a\le a-1$, for all $a>0$, we have

\[
\text{D}\left(f_{0},f\left(\cdot;\bm{\vartheta}_{0}^{\left(g\right)}\right)\right)=\int_{\mathbb{X}}f_{0}\left(\bm{x}\right)\log\left\{ \frac{f_{0}\left(\bm{x}\right)}{f\left(\bm{x};\bm{\vartheta}_{0}^{\left(g\right)}\right)}\right\} \text{d}\bm{x}\le\int_{\mathbb{X}}f_{0}\left(\bm{x}\right)\left\{ \frac{f_{0}\left(\bm{x}\right)}{f\left(\bm{x};\bm{\vartheta}_{0}^{\left(g\right)}\right)}-1\right\} \text{d}\bm{x}\le0\text{,}
\]
which implies that $f\left(\cdot;\bm{\vartheta}_{0}^{\left(g\right)}\right)=f_{0}$,
by (A1) and the definition of the Kullback--Leibler divergence (cf.
\citealp[Lem. 1]{Leroux1992}). Thus, we have the contradiction that
$f_{0}\in\mathcal{M}_{g}$, and hence 
\[
\text{E}_{f_{0}}\log f\left(\bm{X};\bm{\vartheta}_{0}^{\left(g\right)}\right)-\text{E}_{f_{0}}\log f\left(\bm{X};\bm{\vartheta}_{0}^{\left(g+l_{g}\right)}\right)<0\text{,}
\]
as required.

\subsection*{Proof of Corollary \ref{cor: Consistent model selection}}

It suffices to show that for each $\epsilon>0$, there exists a $N\left(\epsilon\right)\in\mathbb{N}$,
such that for all $n_{1},n_{2}\ge N\left(\epsilon\right)$, we have
for any $f_{0}\in\mathcal{M}$:
\[
\mathrm{Pr}_{f_{0}}\left(g_{0}=\hat{g}_{n}\right)\ge1-\epsilon\text{.}
\]
Firstly, using the form of the sequential testing procedure, we can
write
\begin{align*}
\mathrm{Pr}_{f_{0}}\left(g_{0}=\hat{g}_{n}\right) & =\mathrm{Pr}_{f_{0}}\left(\left(\bigcap_{g=1}^{g_{0}-1}\left\{ R_{g}^{n}=1\right\} \right)\cap\left\{ R_{g_{0}}^{n}=0\right\} \right)\\
 & =1-\mathrm{Pr}_{f_{0}}\left(\left(\bigcup_{g=1}^{g_{0}-1}\left\{ R_{g}^{n}=0\right\} \right)\cup\left\{ R_{g_{0}}^{n}=1\right\} \right)\\
 & \ge1-\sum_{g=1}^{g_{0}-1}\mathrm{Pr}_{f_{0}}\left(R_{g}^{n}=0\right)-\mathrm{Pr}_{f_{0}}\left(R_{g_{0}}^{n}=1\right)\text{.}
\end{align*}
By $n_{1}^{-1}\log\alpha_{n}\rightarrow0$ and Theorem \ref{thm: individual test consistency},
and by $\alpha_{n}\rightarrow0$ and Theorem \ref{thm: type 1 control},
we have for any $\delta>0$, there exist $N_{g_{0}}\left(\delta\right)\in\mathbb{N}$,
such that
\[
\mathrm{Pr}_{f_{0}}\left(R_{g}^{n}=0\right)\le\delta\text{,}
\]
and
\[
\mathrm{Pr}_{f_{0}}\left(R_{g_{0}}^{n}=1\right)\le\delta\text{,}
\]
for all $n_{1},n_{2}\ge N_{g_{0}}\left(\delta\right)$ and $g\in\left[g_{0}-1\right]$.
Thus, setting $\epsilon=g_{0}\delta$ and $N\left(\epsilon\right)=\max_{g\in\left[g_{0}\right]}N_{g}\left(\delta\right)$,
we have
\begin{align*}
\mathrm{Pr}_{f_{0}}\left(g_{0}=\hat{g}_{n}\right) & \ge1-\left(g_{0}-1\right)\delta-\delta\\
 & =1-g_{0}\delta=1-\epsilon\text{,}
\end{align*}
as required.

\bibliographystyle{plainnat}
\bibliography{20210119_MASTERBIB}

\end{document}